\documentclass[a4paper,fleqn,usenatbib]{mnras}

\usepackage{newtxtext,newtxmath}

\usepackage[T1]{fontenc}
\usepackage{ae,aecompl}

\usepackage{graphicx}	
\usepackage{amsmath}	
\usepackage{wasysym}

\def\mrpd{\hbox{mrad\,d$^{-1}$}}

\def\chisqr{\hbox{$\chi^2_{\rm r}$}}
\def\msun{\hbox{${\rm M}_{\odot}$}}
\def\mjup{\hbox{${\rm M}_{\jupiter}$}}
\def\me{\hbox{${\rm M}_{\oplus}$}}

\def\mspy{\hbox{${\rm M}_{\odot}$\,yr$^{-1}$}}
\def\rsun{\hbox{${\rm R}_{\odot}$}}
\def\lsun{\hbox{${\rm L}_{\odot}$}}
\def\rcor{\hbox{$r_{\rm cor}$}}
\def\rmag{\hbox{$r_{\rm mag}$}}
\def\mstar{\hbox{$M_{\star}$}}
\def\rstar{\hbox{$R_{\star}$}}
\def\lstar{\hbox{$L_{\star}$}}
\def\teff{\hbox{$T_{\rm eff}$}}
\def\logg{\hbox{$\log g$}}

\def\vD{\hbox{$v_{\rm D}$}}

\def\ms{\hbox{m\,s$^{-1}$}}

\def\kms{\hbox{km\,s$^{-1}$}}

\def\vsini{\hbox{$v \sin i$}}

\def\mic{\hbox{$\mu$m}}

\def\emr{}
\def\Bl{\hbox{$B_{\rm \ell}$}}
\def\Bd{\hbox{$B_{\rm d}$}}

\def\degr{\hbox{$^\circ$}}

\def\Mdot{\hbox{$\dot{M}$}}

\def\Prot{\hbox{$P_{\rm rot}$}}

\newcommand{\caii}{\hbox{Ca$\;${\sc ii}}}

\newcommand{\hei}{\hbox{He$\;${\sc i}}}

\newcommand{\pab}{\hbox{Pa${\beta}$}}
\newcommand{\brg}{\hbox{Br${\gamma}$}}

\title[Magnetic field, accretion and planets of TW~Hya]{SPIRou spectropolarimetry of the T~Tauri star TW~Hydrae: magnetic fields, accretion and planets} 
\author[J.-F.~Donati et al.]{J.-F.~Donati$^{1}$\thanks{E-mail: jean-francois.donati@irap.omp.eu},
           P.I.~Cristofari$^{1,2}$, L.T.~Lehmann$^{1}$, C.~Moutou$^1$, S.H.P.~Alencar$^{3}$,  
\newauthor J.~Bouvier$^{4}$, L.~Arnold$^{5}$, X.~Delfosse$^{4}$, E.~Artigau$^{6}$, N.~Cook$^{6}$, \'A.~K\'osp\'al$^{7,8}$,
\newauthor F.~M\'enard$^{4}$, C.~Baruteau$^{1}$, M.~Takami$^{9}$, S.~Cabrit$^{10}$, G.~H\'ebrard$^{11}$, R.~Doyon$^{6}$ 
\newauthor and the SPIRou science team 
\vspace{1mm}
\\ 
$^1$ Universit\'e de Toulouse, CNRS, IRAP, 14 avenue Belin, 31400 Toulouse, France \\ 
$^2$ Center for Astrophysics, Harvard \& Smithsonian, 60 Garden street, Cambridge, MA 02138, United States \\
$^3$ Universit\'e Grenoble Alpes, CNRS, IPAG, 38000 Grenoble, France \\ 
$^4$ Departamento de F\'{\i}sica -- ICEx -- UFMG, Av. Ant\^onio Carlos, 6627, 30270-901 Belo Horizonte, MG, Brazil\\  
$^5$ Canada-France-Hawaii Telescope, 65-1238 Mamalahoa Hwy., Kamuela, HI 96743, USA \\ 
$^6$ Universit\'e de Montr\'eal, D\'epartement de Physique, IREX, Montr\'eal, QC H3C 3J7, Canada \\ 
$^7$ Konkoly Observatory, HUN-REN Research Centre for Astronomy and Earth Sciences, Konkoly-Thege Mikl\'os \'ut 15-17, 1121 Budapest, Hungary \\ 
$^8$ Institute of Physics and Astronomy, ELTE E\"otv\"os Lor\'and University, P\'azm\'any P\'eter s\'et\'any 1/A, 1117 Budapest, Hungary \\ 
$^9$ Institute of Astronomy and Astrophysics, Academia Sinica, Roosevelt Rd, Taipei 10617, Taiwan \\
$^{10}$ Observatoire de Paris, CNRS, LERMA, Sorbonne Universit\'e, 61 avenue de l'Observatoire, 75014 Paris, France \\ 
$^{11}$ Institut d'Astrophysique de Paris, CNRS, Sorbonne Univ., 98 bis bd Arago, 75014 Paris, France  
}

\date{Submitted 2024 xxx -- Accepted 2024 xxx } 

\pubyear{2023}

\begin{document}

\label{firstpage}
\pagerange{\pageref{firstpage}--\pageref{lastpage}}
\maketitle

\begin{abstract}
In this paper we report near-infrared observations of the classical T~Tauri star TW~Hya with the SPIRou high-resolution spectropolarimeter and velocimeter at the 3.6-m 
Canada-France-Hawaii Telescope in 2019, 2020, 2021 and 2022.  By applying Least-Squares Deconvolution (LSD) to our circularly polarized spectra, we derived 
longitudinal fields that vary from year to year from --200 to +100~G, and exhibit low-level modulation on the 3.6~d rotation period of TW~Hya, despite the star being 
viewed almost pole-on.  We then used Zeeman-Doppler Imaging to invert our sets of unpolarized and circularly-polarized LSD profiles into brightness and magnetic maps 
of TW~Hya in all 4 seasons, and obtain that the large-scale field of this T~Tauri star mainly consists of a 1.0--1.2~kG dipole tilted at about 20\degr\ to the rotation axis, 
whereas the small-scale field reaches strengths of up to 3-4~kG.  We find that the large-scale field is strong enough to allow TW~Hya to accrete material from the 
disc on the polar regions at the stellar surface in a more or less geometrically stable accretion pattern, but not to succeed in spinning down the star.  
We also report the discovery of a radial velocity signal of semi-amplitude $11.1^{+3.3}_{-2.6}$~\ms\ (detected at 4.3$\sigma$) at a period of 8.3~d in the spectrum of TW~Hya, 
whose origin may be attributed to either a non-axisymmetric density structure in the inner accretion disc, or to a $0.55^{+0.17}_{-0.13}$~\mjup\ candidate close-in planet 
(if orbiting in the disc plane), at an orbital distance of $0.075\pm0.001$~au.  
\end{abstract}




\section{Introduction}
\label{sec:int}

It is now well established, through documented observations collected with various instruments over the last decades, that stars and their planets form at the same time, 
following the collapse of large gravitationally unstable turbulent molecular clouds.  The cloud collapse results in an accretion disc that feeds the central protostar from 
its inner regions, and where growing protoplanets form from merging planetesimals, giving birth, in the outer disc regions, to massive planets that can migrate inwards into 
close-in hot Jupiters \citep[e.g.,][]{Baruteau14}.  Magnetic fields play a key role in this process in many different ways, e.g., by hampering fragmentation within the disc 
\citep{Hennebelle08b}, by evacuating the central regions of the accretion disc and forcing disc material to flow along discrete magnetospheric funnels linking the inner disc 
to the surface of the host star \citep[e.g.,][]{Bouvier14}, by extracting angular momentum outwards through outflows and jets and forcing the central star to slow down through 
star / disc interactions \citep[e.g.,][]{Romanova02,Zanni13}, or by making inward-migrating giant planets pile-up at the outer edge of the magnetosphere and thereby saving 
them from falling into their host stars \citep[e.g.,][]{Lin96,Romanova06,Mulders15}.  

T~Tauri stars (TTSs), and in particular classical TTSs (cTTSs\footnote{{\emr All abbreviations used in the paper are listed in Sec.~\ref{sec:appD} for easier reference.}}) 
that still accrete material from their accretion discs, are ideal objects to investigate these critical 
steps of star / planet formation, and to yield observational constraints on the complex physics at work, especially in the inner regions that drive star / disc / planet interactions 
and where the most energetic phenomena take place.  However, although accretion discs of cTTSs are presumably actively forming planets given the radial structuring of their density 
profiles \citep[e.g.,][]{Clarke18}, detecting planets around cTTSs has proven quite complex, either because of limited performances in angular resolution and contrast for direct imaging searches, 
or due to the extreme level of intrinsic variability cTTSs are subject to \citep[e.g.,][]{Cody14,Sousa16} that drastically limit the precision of indirect velocimetric measurements.  
So far, distant planets have only been reliably detected around a single cTTSs \citep{Keppler18,Haffert19}, and claims of close-in hot Jupiters 
detected through velocimetry often turned out to be false positives attributable to activity \citep{Setiawan08,Huelamo08,JohnsKrull16,Donati20b}.  Moreover, planet detection 
through transit events looks hopeless given the small size of expected transits, even for massive planets, with respect to the huge intrinsic variability induced by ongoing 
accretion processes between the disc and the central star.  

Located at the heart of the TW~Hydra (TWA) association about 60~pc away from the Sun, TW~Hya, aged about 8--10~Myr \citep[e.g.,][]{Torres08,Luhman23}, is the closest and most 
studied cTTS \citep[e.g.,][]{Rucinski83,Kastner02,Herczeg23} that 
hosts a large and massive accretion disc which survived longer than the typical disc dissipation timescale \citep[of about 3~Myr, e.g.,][]{Kraus12}, and features rings and 
gaps suggesting ongoing planet formation \citep{vanBoekel17}.  TW~Hya is also known to harbour a strong large-scale magnetic field \citep{Yang07,Donati11} that interacts with the 
inner accretion disc, as well as small-scale fields locally reaching up to several kG \citep[e.g.,][]{Sokal18,Lavail19,Lopez-Valdivia21}.  TW~Hya is thus an ideal laboratory to scrutinize 
magnetospheric accretion processes and more generally star / planet formation and interactions in the inner discs of cTTSs.  

In this paper, we report extended observations of TW~Hya with the near-infrared (nIR) high-resolution cryogenic spectropolarimeter / velocimeter SPIRou installed at the 
Cassegrain focus of the Canada-France-Hawaii Telescope (CFHT), carried out as a monitoring program over 4 consecutive observing seasons (2019, 2020, 2021 and 2022).  
After describing the observational material we collected (in Sec.~\ref{sec:obs}) and briefly summarizing the latest estimates for the main stellar atmospheric parameters
(in Sec.~\ref{sec:par}), we present our measurements of the large-scale magnetic field of TW~Hya (in Sec.~\ref{sec:bl}) and the magnetic modeling that we derive from 
our spectropolarimetric observations using tomographic imaging techniques (in Sec.~\ref{sec:zdi}).  We then outline our radial velocity (RV) measurements of TW~Hya and their 
modeling in a Bayesian framework (in Sec.~\ref{sec:rvs}), and describe the characteristics of the nIR emission lines traditionally probing accretion in cTTSs, as well as their 
temporal behaviour (in Sec.~\ref{sec:eml}).  We finally conclude our study and discuss its implications for our understanding of star / planet formation in magnetized cTTSs 
(in Sec.~\ref{sec:dis}).

\section{SPIRou observations}
\label{sec:obs}

TW~Hya was observed over several successive seasons with the SPIRou nIR spectropolarimeter / high-precision velocimeter \citep{Donati20} at CFHT, first within 
the SPIRou Legacy Survey (SLS) in 2019, 2020 and 2021, then within the PI program of Lisa Lehmann in 2022 (run IDs 22AF14 and 22AF96).  SPIRou  
collects unpolarized and polarized stellar spectra, covering a wavelength interval of 0.95--2.50~\mic\ at a resolving power of 70\,000 in a single exposure.  
For the present study, we concentrated on circularly polarized (Stokes $V$) and unpolarized (Stokes $I$) spectra of TW~Hya only.  
Each polarization observation consists of a sequence of 4 sub-exposures, associated with different azimuths of the Fresnel rhomb retarders in order to remove 
systematics in polarization spectra \citep[to first order, see, e.g.,][]{Donati97b}.  Each sequence yields one Stokes $I$ and one Stokes $V$ spectrum, 
as well as a null polarization check (called $N$) allowing one to diagnose potential instrumental or data reduction issues.  

A total of 84 polarization sequences of TW~Hya were collected in 4 main seasons, 11 in 2019 (April), 14 in 2020 (February to May), 30 in 2021 (February to May), 
and 29 in 2022 (March to May).  A single polarization sequence was recorded in most nights;  however, in a few cases (on 2021 March 28, April 26, April 28 and 
2022 March 23), a second sequence was collected when data quality was lower than usual in the first one.  Two spectra were discarded due to very low signal to 
noise ratios (SNRs), one in 2021 (April 28) and another one in 2022 (May 18).  It finally yielded a total of 82 usable Stokes $I$, $V$ and $N$ spectra of TW~Hya, 
with  11, 14, 29 and 28 of them in 2019, 2020, 2021 and 2022 respectively, spanning in each case time slots of 12, 100, 70 and 70~d, and altogether covering a 
temporal window of 1131~d.  The full log of our observations is provided in Table~\ref{tab:log} of Appendix~\ref{sec:appA}.   

Our SPIRou spectra were processed with \texttt{Libre ESpRIT}, the nominal reduction pipeline of ESPaDOnS at CFHT, optimized for spectropolarimetry and adapted 
for SPIRou \citep{Donati20}.  
Least-Squares Deconvolution \citep[LSD,][]{Donati97b} was then applied to all reduced spectra, using a line mask constructed from the VALD-3 database 
\citep{Ryabchikova15} for an effective temperature \teff=4000~K and a logarithmic surface gravity \logg=4.0 adapted to TW~Hya (see Sec~\ref{sec:par}).  
Atomic lines of relative depth larger than 10 percent were selected, for a total of $\simeq$1300 lines, featuring an average wavelength and Land\'e factor of 
1750~nm and 1.2 respectively.  The noise levels $\sigma_V$ in the resulting Stokes $V$ LSD profiles range from 1.1 to 3.8 (median 1.6, in units of $10^{-4} I_c$ 
where $I_c$ denotes the continuum intensity).  We also applied LSD with a mask containing the CO lines of the CO bandhead (at 2.3~\mic) only, to obtain veiling 
estimates in the K band, in addition to those for the whole spectrum derived from LSD profiles of atomic lines (see Sec.~\ref{sec:par}).  
Phases and rotation cycles were derived assuming a rotation period of $\Prot=3.606$~d (see Sec.~\ref{sec:bl}) 
and counting from an arbitrary starting BJD0 of 2458488.5 (i.e., just prior to our first SPIRou observation).  

Our data were also processed with \texttt{APERO} (version 0.7.288), the nominal SPIRou reduction pipeline \citep{Cook22} optimized for RV precision.  The 
reduced spectra were first analyzed for Zeeman broadening (see Sec.~\ref{sec:bl}), then for RVs (see Sec.~\ref{sec:rvs}) with the line-by-line (LBL) technique 
\citep[version 0.63,][]{Artigau22}.  It yielded 79 nightly RVs \citep[corrected from instrumental drifts monitored with the SPIRou RV reference module,][]{Donati20} 
and associated error bars (median 2.2~\ms), listed in Table~\ref{tab:log}.

\section{Fundamental parameters of TW~Hya}
\label{sec:par}

In this section we recall the main parameters of TW~Hya from the literature, including our previous study from ESPaDOnS spectra \citep{Donati11}.  
TW~Hya is a cTTS located at a distance of $59.96^{+0.37}_{-0.11}$~pc from the Sun \citep{Gaia23}, in the TWA association, and hosts the protoplanetary disc 
closest to the Solar System.  According to Gaia again, TW~Hya features a photospheric temperature of $\teff=3850\pm10$~K, a logarithmic surface gravity of 
$\logg=4.05\pm0.01$~dex (cgs units) and a metallicity relative to the Sun of ${\rm [M/H]}=-0.50\pm0.05$~dex.  We suspect that, despite the small 
error bars, these estimates, and in particular ${\rm [M/H]}$, are likely affected by surface spots induced by magnetic activity and by intrinsic variability 
caused by accretion from the surrounding disc \citep[e.g.,][]{Herczeg23}.  Although \citet{Herczeg14} also suggest, based on low-resolution spectra, 
that $\teff\simeq3810$~K (corresponding to a M0.5 spectral type) looks adequate, several studies involving high-resolution spectra rather conclude 
that \teff\ is higher and more consistent with a K7 spectral type \citep[e.g.,][]{Torres03,Yang05}.  

Applying our own spectral characterization tool on selected atomic lines in the ESPaDOnS spectra of TW~Hya from our previous study yields 
$\teff=4060\pm50$~K and $\logg=4.2\pm0.1$~dex assuming solar metallicity, as appropriate for nearby young stars \citep[e.g.,][]{Padgett96}.  
When applying ZeeTurbo \citep[a tool specifically developed for SPIRou spectra of M dwarfs,][]{Cristofari22a,Cristofari22b,Cristofari23,Cristofari23b} 
to our SPIRou spectra of TW~Hya, and assuming again solar metallicity, we obtain $\teff=3970\pm50$~K, including the effect of magnetic fields (see 
Sec.~\ref{sec:bl}) but fixing \logg\ to 4.2 and the veiling in the YJH and K bands to 0.20 and 0.25 respectively (as measured, see below) to minimize correlation 
between fitted parameters.  This is slightly cooler than (though still consistent with) our estimate from ESPaDOnS spectra, likely as a result of cool 
magnetic spots at the surface of TW~Hya \citep[][see also the following sections]{Huelamo08,Donati11} that affect nIR lines more than optical ones (the lower brightness contrast 
between cool spots and the photosphere in the nIR implying a larger relative contribution of these spots to nIR lines).  
This is why we did not include the CO lines of the CO bandhead in 
the fit, as these lines get stronger with decreasing temperature and thereby further bias the determination of \teff\ towards spot temperatures.  
Our measurements are consistent with those of another study from high-resolution nIR spectra of TW~Hya \citep[][also taking into account magnetic fields 
and assuming solar metallicity]{Sokal18}, yielding $\teff=3800\pm100$~K and $\logg=4.2\pm0.1$~dex.  The agreement is best for \logg\ and less so for \teff, 
presumably for the reason already outlined above, i.e., the presence of cool star spots affecting temperature determination \citep[see also][]{Gully17}.  

In the following, we use the estimates derived from our optical data, less affected by cool surface spots.  
Comparing with \logg\ vs \teff\ synthetic tracks from the evolutionary models of \citet{Baraffe15} yields for TW~Hya a mass of $\mstar=0.80\pm0.05$~\msun\ 
\citep[in good agreement with the dynamical mass derived from ALMA observations of the almost pole-on accretion disc, equal to $0.81\pm0.16$~\msun,][]{Teague19}, 
an age of $7.5\pm2.5$~Myr, a radius of $\rstar=1.16\pm0.13$~\rsun\ consistent with interferometric measurements \citep[$1.29\pm0.19$~\rsun,][]{Gravity20}, 
and a logarithmic luminosity with respect to the Sun of $\log(\lstar/\lsun)=-0.48\pm0.10$.  
These evolution models also predict that TW~Hya already started to develop a radiative core of mass $0.2\pm0.1$~\msun.  

The rotation period of TW~Hya was unambiguously determined to be about 3.6~d from the reported periodic changes in RV, that were first erroneously 
attributed to the reflex motion of a putative massive close-in planet \citep{Setiawan08} then to magnetic activity inducing rotational modulation 
\citep{Huelamo08,Donati11,Sicilia23}.  Our new spectropolarimetric data confirm this value, with the line-of-sight component of the large-scale 
magnetic field integrated over the visible stellar hemisphere, called longitudinal magnetic field and denoted \Bl, steadily varying with a period 
of $3.606\pm0.010$~d throughout the 1131~d of our observations (see Sec.~\ref{sec:bl}).  
The corresponding corotation radius, at which the Keplerian rotation rate equals that 
at the surface of the star, is equal to $\rcor=0.043\pm0.001$~au ($7.9\pm0.3$~\rstar).  
We note that photometric observations of TW~Hya \citep[e.g.,][]{Rucinski08, Siwak14, Siwak18}, including the most recent ones collected with TESS in March 2019, 2021 and 2023 
(each lasting about 25~d, and the first two contemporaneous with our SPIRou spectra), rarely exhibit rotational modulation, but rather a spectrum of unstable periods of order 
a few days likely probing intrinsic variability triggered by unsteady accretion from the inner regions of the accretion disc.  For instance, 
Fig.~\ref{fig:tes} shows stacked periodograms of the March 2019 and 2021 binned TESS light curve, that exhibit transient periodic signals, with only little power at 
the rotation period \citep[see also][]{Sicilia23}.  

Given this rotation period and the stellar radius derived above ($1.16\pm0.13$~\rsun), 
we can conclude that the line-of-sight projected equatorial rotation velocity \vsini\ of TW~Hya must be small if the rotation axis of the star is 
close to the line of sight, as is that of the accretion disc \citep[$i=5.8^{+4.0}_{-1.7}$ for the disc,][]{Teague19}.  For instance, $\vsini=5.8$~\kms\ 
\citep[][]{Sokal18} or $\vsini=8.4$~\kms\ \citep[][]{Lopez-Valdivia21} would imply inclination angles of the stellar axis to the line of sight of 
$i\simeq20$\degr\ and 30\degr\ respectively, i.e., significantly larger than that of the accretion disc.  
We adopt here a value of $\vsini=3\pm1$~\kms\ corresponding to an inclination of the stellar rotation axis $i\simeq10$\degr, 
slightly larger than, though still consistent with, that of the outer disc.  
In addition to yielding a better match to the observed Stokes $I$ profiles than the larger estimates mentioned above (see Sec.~\ref{sec:zdi}), 
this \vsini\ is more consistent with our previous measurement \citep[i.e., $\vsini=4\pm1$~\kms,][]{Donati11}, with that of other independent studies 
\citep[e.g.,][]{Sicilia23}, and with the inclination of the outer disc.  

As mentioned above, the accretion disc of TW~Hya, extending to a few hundred au's, features dusty rings and gaps 
\citep[including one at 1~au from the star,][]{Andrews16,Nomura16,vanBoekel17,Nomura21} as well as 
spiral structures \citep{Teague19}, possibly tracing the presence of planets forming and migrating throughout the disc.  
The detection of moving shadows at the surface of the outer disc \citep{Debes17} also suggests that the inner disc regions 
may not be coplanar with the outer ones \citep{Teague22,Debes23}, hence potentially supporting that the rotation axis of the central star is slightly 
misaligned with that of the disc, and / or that planets are indeed present in the innermost regions of the disc.  
Although low, mass accretion is still observed at the surface of the star \citep[at an average rate of $\log\Mdot=-8.65$~\mspy, and with values ranging from 
--9.2 to --8.2 over time,][see also Sec.~\ref{sec:eml}]{Herczeg23,Sousa23}, proceeding through 
discrete accretion funnels linking the inner disc to the stellar surface \citep{Donati11}, and generates only low veiling at nIR wavelengths \citep{Sousa23}.  
Interferometric observations indicate that the inner disc edge from which gas is accreted is located at $0.021\pm0.001$~au, i.e., well within the corotation 
radius, whereas dust is only present from $0.039\pm0.001$~au outwards \citep{Gravity20}, i.e., from about the corotation radius and beyond.   

To double check veiling estimates, we used our LSD Stokes $I$ profiles of TW~Hya, both for atomic lines over the whole spectrum and for the CO lines of the CO 
bandhead in the K band, and compared them with the corresponding median LSD Stokes $I$ profiles of the weak-line TTS TWA~25, also observed with SPIRou and that we take as 
a reference for an unveiled spectrum of similar spectral type \citep[as in][]{Sousa23}.  We find that the median veiling is about 20\% over the whole spectrum, 
and only marginally larger (25\%) in the K band (CO bandhead), slightly stronger than (though still consistent with) the estimates of \citet{Sousa23} and 
in agreement with measurements derived from older nIR spectra of TW~Hya \citep{Lavail19}.  Besides, we obtain that the median veiling per season is about the same 
(within a few \%) for all four seasons.  We also confirm that veiling in individual spectra is about 2.5$\times$ more variable with time in the K band (about 15\% 
rms about median veiling) than in the rest of the spectrum, again qualitatively consistent with \citet{Sousa23}.  

Table~\ref{tab:par} summarizes the main parameters of TW~Hya used in, or derived from, the present study.

\begin{table}
\caption[]{Parameters of TW~Hya used in / derived from our study} 
\scalebox{0.95}{\hspace{-4mm}
\begin{tabular}{ccc}
\hline
distance (pc)        & $59.96^{+0.37}_{-0.11}$  & \citet{Gaia23} \\
$\log(\lstar/\lsun)$ & $-0.48\pm0.10$  & \\ 
\teff\ (K)           & $4050\pm50$     & from ESPaDOnS spectra \\
\logg\ (dex)         & $4.2\pm0.1$     & from ESPaDOnS spectra \\ 
\mstar\ (\msun)      & $0.80\pm0.05$   & using \citet{Baraffe15} \\
\rstar\ (\rsun)      & $1.16\pm0.13$   & using \citet{Baraffe15} \\
age (Myr)            & $7.5\pm2.5$     & using \citet{Baraffe15} \\ 
\Prot\ (d)           & $3.606$         & period used to phase data \\ 
\Prot\ (d)           & $3.606\pm0.010$ & period from \Bl\ data \\ 
\vsini\ (\kms)       & $3\pm1$         & \citet{Donati11} \\ 
<$B$> (kG)           & $3.60\pm0.04$   & on median spectrum \\
$a_0$ (\%)           & $7\pm2$         & on median spectrum \\ 
$i$ (\degr), star    & $10$            & used for ZDI \\ 
$i$ (\degr), disc    & $5.8^{+4.0}_{-1.7}$ & \citet{Teague19} \\ 
\rcor\ (au)          & $0.043\pm0.001$  & from \mstar\ and \Prot \\ 
\rcor\ (\rstar)      & $7.9\pm0.3$      &  \\ 
$\log\Mdot$ (\mspy)  & $-8.65\pm0.22$   & \citet{Herczeg23} \\ 
$\log\Mdot$ (\mspy)  & $-8.72\pm0.22$   & from \pab\ \& \brg \\ 
\hline
\end{tabular}}
\label{tab:par}
\end{table}

\section{The longitudinal field and Zeeman broadening of TW~Hya}
\label{sec:bl}

The next step in our analysis is to derive the longitudinal field \Bl\ of TW~Hya following \citet{Donati97b}, from each of the Stokes $V$ and $I$ LSD profiles 
derived in Sec.~\ref{sec:obs}.  In practice, we computed the first moment of the Stokes $V$ profile and its error bar, whereas the equivalent width of the 
Stokes $I$ LSD profiles is measured through a Gaussian fit.  Stokes $V$ LSD signatures were integrated over a window of $\pm40$~\kms\ in the stellar rest frame, 
given the strong magnetic broadening of line profiles \citep{Yang05,Sokal18,Lavail19}, with the exact integration width having little impact on the result.  We proceeded 
in the same way with the polarization check $N$ to verify that the 
derived pseudo longitudinal field is consistent with 0 within the error bars, i.e., associated with a reduced chi-square \chisqr\ close to unity.  The inferred \Bl\ values 
computed from our 82 Stokes $V$ profiles are listed in Table~\ref{tab:log} and range from $-195$ to 77~G (median $-37$~G) with error bars of 10 to 34~G (median 
15~G), yielding a \chisqr\ (with respect to the $\Bl=0$~G line) equal to 32.3 for $V$ (1.09 for $N$).  This demonstrates that the magnetic field is unambiguously 
detected in the Stokes $V$ signatures of TW~Hya, that no spurious pollution is observed in $N$ and that our analytical error bars are consistent with the observed 
dispersion within 5\%.  Unsurprisingly, our \Bl\ values are an order of magnitude weaker than the small-scale fields estimated from the Zeeman broadening of nIR 
lines \citep[$\simeq$3~kG,][see also below]{Yang05,Sokal18,Lavail19,Lopez-Valdivia21}, as usual for cool active stars harboring small-scale tangled fields whose circular polarization signatures 
mostly cancel out.  

\begin{figure*}
\centerline{\includegraphics[scale=0.6,angle=-90]{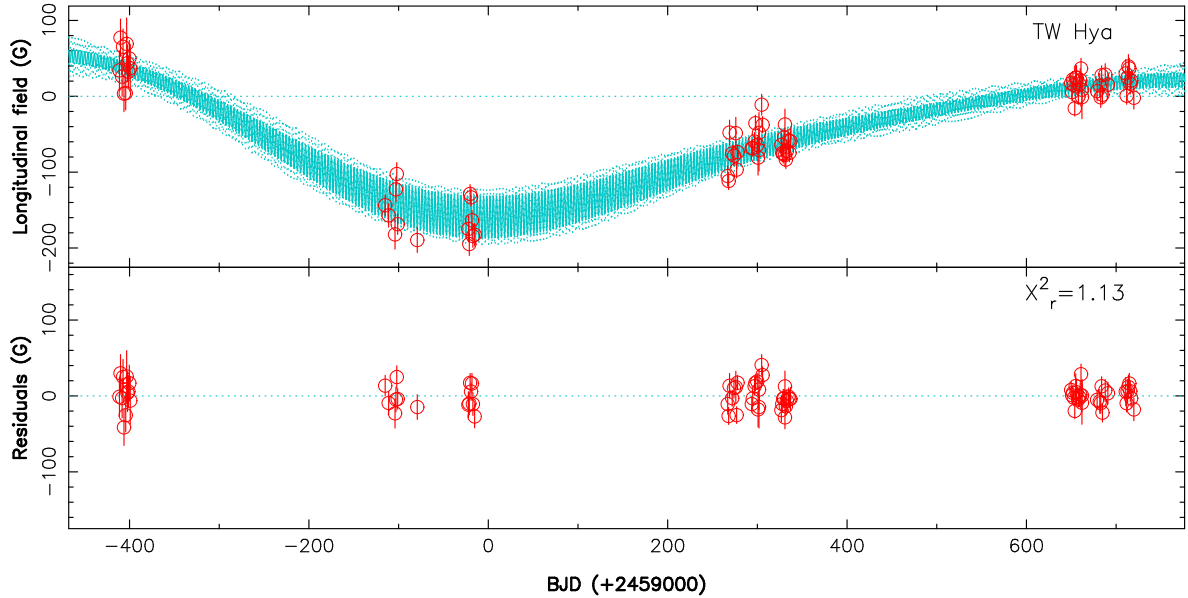}}
\caption[]{Longitudinal magnetic field \Bl\ of TW~Hya (red open circles) as measured with SPIRou throughout our campaign, and QP GPR fit to the data (cyan full line) with 
corresponding 68\% confidence intervals (cyan dotted lines).  The residuals are shown in the bottom panel.   The rms of the residuals is about 15~G ($\chisqr=1.13$), 
consistent with our median error bar (of 15~G as well).  The \chisqr\ with respect to the $\Bl=0$~G line is equal to 32.3.}  
\label{fig:gpb}
\end{figure*}

We then investigate the temporal behaviour of our \Bl\ data, arranged in a vector denoted $\bf y$, using quasi-periodic (QP) Gaussian-Process Regression (GPR), with 
a covariance function $c(t,t')$ of type: 
\begin{eqnarray}
c(t,t') = \theta_1^2 \exp \left( -\frac{(t-t')^2}{2 \theta_3^2} -\frac{\sin^2 \left( \frac{\pi (t-t')}{\theta_2} \right)}{2 \theta_4^2} \right) 
\label{eq:covar}
\end{eqnarray}
where $\theta_1$ is the amplitude (in G) of the Gaussian Process (GP), $\theta_2$ its recurrence period (i.e., \Prot, in d), $\theta_3$ the evolution timescale 
(in d) on which the shape of the \Bl\ modulation changes, and $\theta_4$ a smoothing parameter describing the amount of harmonic complexity needed to describe the data.  
We then select the QP GPR fit that features the highest likelihood $\mathcal{L}$, defined by: 
\begin{eqnarray}
2 \log \mathcal{L} = -n \log(2\pi) - \log|{\bf C+\Sigma+S}| - {\bf y^T} ({\bf C+\Sigma+S})^{-1} {\bf y}
\label{eq:llik}
\end{eqnarray}
where $\bf C$ is the covariance matrix for our 82 epochs, $\bf \Sigma$ the diagonal variance matrix associated with $\bf y$, and ${\bf S}=\theta_5^2 {\bf J}$ ($\bf J$ 
being the identity matrix) the contribution from an 
additional white noise source that we introduce as a fifth hyper-parameter $\theta_5$ (in case our error bars on \Bl\ were underestimated for some reason).  
The hyper-parameter domain is then explored using a Monte-Carlo Markov Chain (MCMC) process, yielding posterior distributions and error bars for all hyper-parameters.  

\begin{table} 
\caption[]{Results of our MCMC modeling of the \Bl\ curve of TW~Hya with QP GPR.  For each hyper-parameter, we list the fitted value, the corresponding error bar 
and the assumed prior.  The knee of the modified Jeffreys prior is set to $\sigma_{B}$, i.e., the median error bar of our \Bl\ measurements (i.e., 15~G). 
We also quote the resulting \chisqr\ and rms of the final GPR fit. }  
\scalebox{0.95}{\hspace{-6mm}
\begin{tabular}{cccc}
\hline
Parameter   & Name & Value & Prior   \\
\hline 
GP amplitude (G)     & $\theta_1$  & $74^{+19}_{-15}$  & mod Jeffreys ($\sigma_{B}$) \\
Rec.\ period (d)     & $\theta_2$  & $3.606\pm0.015$   & Gaussian (3.6, 0.5) \\
Evol.\ timescale (d) & $\theta_3$  & $232^{+64}_{-50}$ & log Gaussian ($\log$ 250, $\log$ 2) \\
Smoothing            & $\theta_4$  & $2.50$            & fixed \\
White noise (G)      & $\theta_5$  & $9.4^{+2.7}_{-2.1}$  & mod Jeffreys ($\sigma_{B}$) \\
\hline
\chisqr              &  \multicolumn{3}{c}{1.13}              \\ 
rms (G)              &  \multicolumn{3}{c}{14.8}              \\ 
\hline 
\end{tabular}}      
\label{tab:gpr}      
\end{table}          

The results of the GPR fit are shown in Fig.~\ref{fig:gpb} whereas the derived hyper-parameters are listed in Table~\ref{tab:gpr}.  The first surprising conclusion 
is that \Bl\ is changing sign from epoch to epoch, being mostly positive in 2019 and 2022 but negative in 2020 and 2021.  As we will see in Sec.~\ref{sec:zdi}, this 
does not reflect an overall polarity switch of the magnetic field, but rather results from small changes of the large-scale topology and / or of the surface brightness distribution 
in an almost pole-on viewing configuration.  This is fairly different from the results of our earlier optical observations with ESPaDOnS, where the longitudinal field in photospheric 
lines was always positive, reaching several hundred G, and that from accretion lines \citep[\caii\ infrared triplet and \hei\ $D_3$ lines in particular,][]{Donati11} was always negative.  
What we detect with SPIRou is in between, reflecting that magnetic regions are spatially associated with dark surface features, with a spot-to-photosphere brightness 
contrast that is smaller in the infrared than in the optical (see Sec.~\ref{sec:zdi}) as was the case for CI~Tau \citep{Donati24}.  

Our second main result is that we are able to detect rotational modulation of \Bl, especially in 2020, despite the amplitude of the modulation being small as a 
result of the close to pole-on viewing configuration.  The rotation period we measure is equal to $\Prot=3.606\pm0.015$~d, slightly larger than that derived from optical 
RV data \citep[i.e., $3.5683\pm0.0002$~d, ][]{Huelamo08}.  This suggests that weak differential rotation is present at the surface of TW~Hya (at a level of only a few 
\mrpd), as further discussed in Sec.~\ref{sec:rvs}.  We also find that the rotational modulation of the \Bl\ curve is simple enough for the GPR fit to yield a smoothing 
parameter that is large and weakly constrained by the data, which we thus fix at its optimal value ($\theta_4=2.5$).  The evolution timescale ($\theta_3=232^{+64}_{-50}$~d) 
is about 3$\times$ longer than that of the more evolved young active star AU~Mic, whose \Bl\ curve is also more complex \citep{Donati23}.  We finally outline that \Bl\ can 
almost be fitted down to the noise level ($\chisqr=1.13$) and thus that the additional white noise term $\theta_5$ is only slightly larger than zero.  

\begin{figure}
\includegraphics[scale=0.32,bb=20 50 800 500]{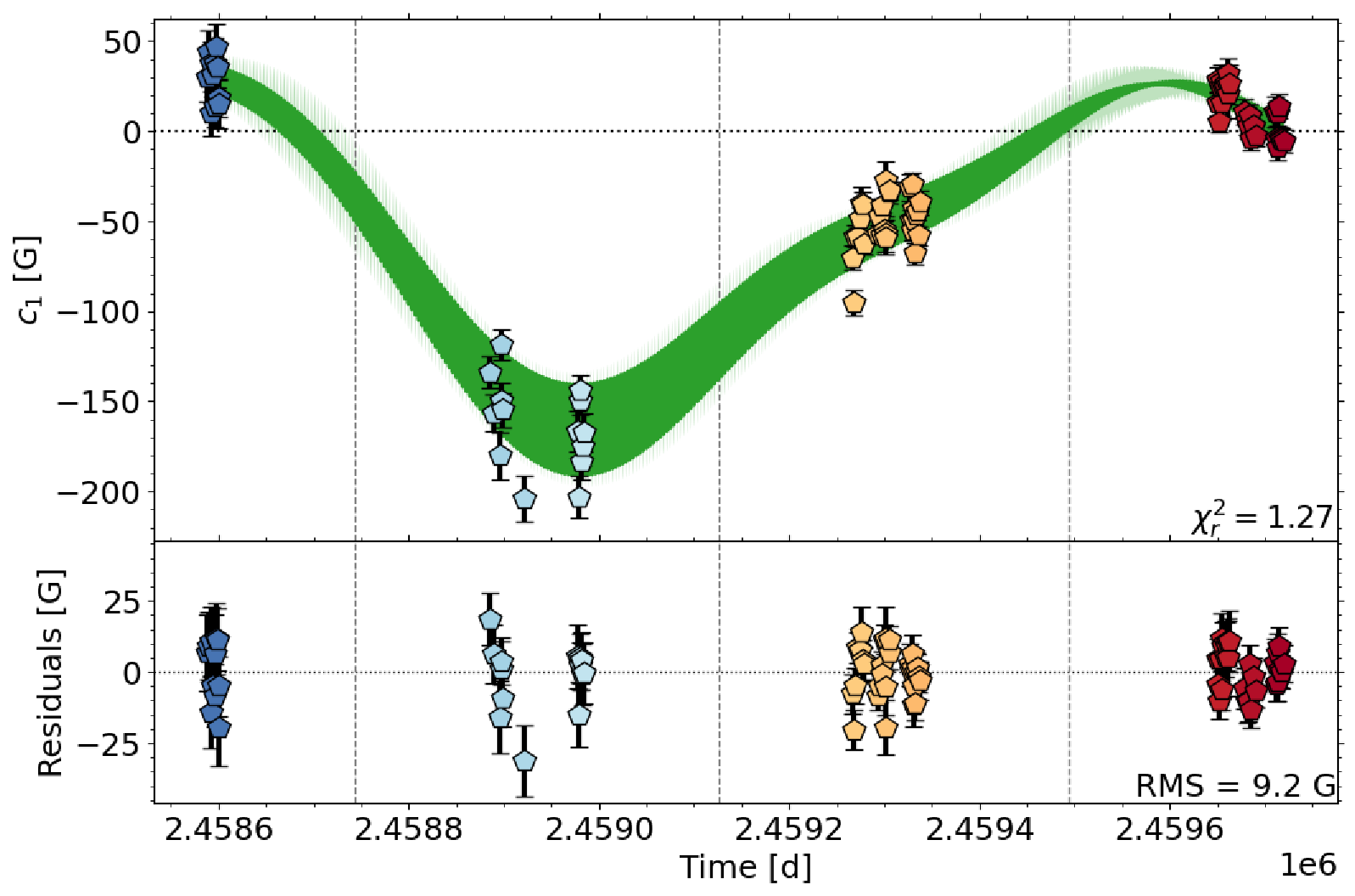}
\caption[]{Same as Fig.~\ref{fig:gpb} for the $c_1$ coefficient of the first PCA eigenvector (scaled and shifted to match \Bl).  The GPR fit and associated error bars are 
now shown in green.  The rms of the residuals is about 9~G.} 
\label{fig:pcac1}
\end{figure}

We also carried out a Principal Component Analysis (PCA) of our Stokes $V$ profiles \citep[as outlined in][]{Lehmann22,Lehmann24} and find that the first PCA eigenvector is 
capable of reproducing most of the observed variations of the mean-subtracted Stokes $V$ profiles (see Fig.~\ref{fig:pca}).  The second PCA eigenvector, encoding wavelength 
shifts of the Stokes $V$ profiles (given its shape that mimics the derivative of the first eigenvector), is only marginally required, indicating that the parent magnetic 
regions do not travel much throughout the line profile, i.e., are located at high latitudes.  We also find that the mean Stokes $V$ profile is antisymmetric with respect to 
the line center and dominates over the mean-subtracted profiles, i.e., that the large-scale field is mainly poloidal and axisymmetric, as expected from the nearly pole-on 
configuration of TW~Hya that renders us almost insensitive to axisymmetric toroidal fields at the surface of the star (nearly perpendicular to the line of sight).  The 
$c_1$ coefficient associated with the first PCA eigenvector \citep[scaled and shifted to match \Bl, as in][]{Lehmann24} exhibits a time dependence very similar to that of \Bl\ 
(see Fig.~\ref{fig:pcac1}), with little to no rotational modulation depending on the season (see Fig.~\ref{fig:pca}), typical of a simple poloidal field only slightly tilted 
to the rotation axis.  These preliminary conclusions are confirmed with the full magnetic modeling of TW~Hya presented in Sec.~\ref{sec:zdi}.  

\begin{figure}
\centerline{\includegraphics[scale=0.6,bb=20 20 380 390]{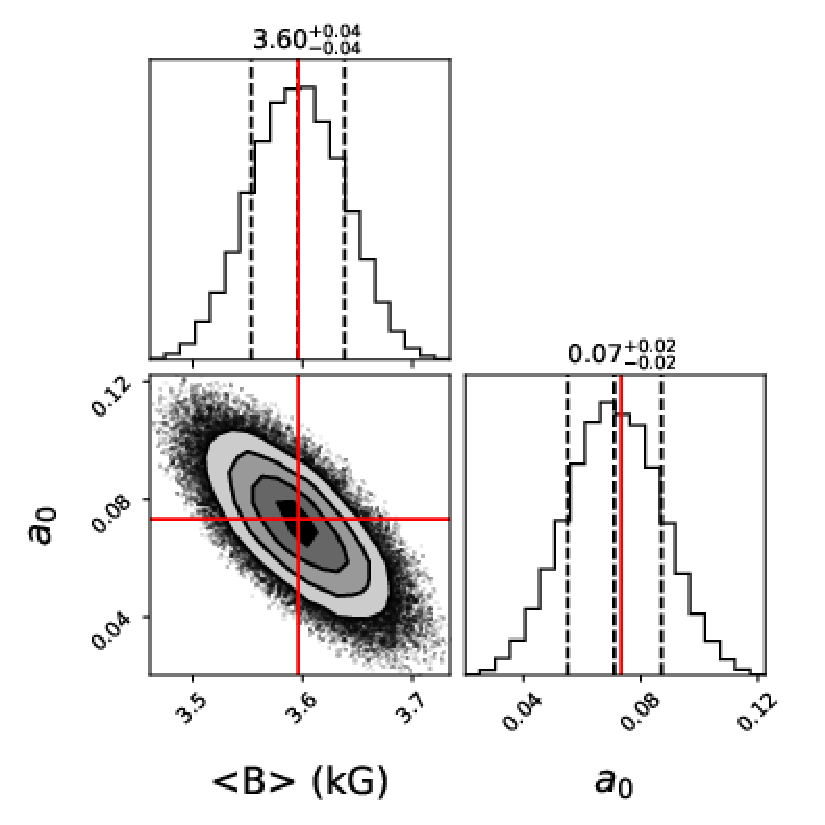}}
\caption[]{Magnetic parameters of TW~Hya, derived by fitting our median SPIRou spectrum using the atmospheric modeling approach of \citet{Cristofari23},
which incorporates magnetic fields as well as a MCMC process to determine optimal parameters and their error bars.  We find that TW~Hya hosts a small-scale
magnetic field of <$B$>~$=3.60\pm0.04$~kG, with the relative area of non-magnetic regions being $a_0=7\pm2$\%.  }
\label{fig:bmag}
\end{figure}

{\emr Looking at the Zeeman broadening of atomic and molecular lines with ZeeTurbo \citep{Cristofari23,Cristofari23b} applied to our median spectrum of TW~Hya, we find 
that 4 components, associated with small-scale magnetic fields of strengths 0, 2, 4 and 6~kG, and respective filling factors $a_0=7\pm2$\%, $a_2=40\pm3$\%, $a_4=20\pm3$\% and 
$a_6=33\pm2$\% of the visible stellar surface, are needed to obtain a good fit, yielding an overall small-scale field measurement of <$B$>~$=3.60\pm0.04$~kG (see Fig.~\ref{fig:bmag}).  
Our estimate of the small-scale field of TW~Hya from SPIRou spectra is consistent with those derived in previous studies \citep{Yang05,Sokal18,Lavail19,Lopez-Valdivia21}, given 
the expected temporal variability.  On the timescale of our observations, we find only marginal variations of the small-scale field (typically 0.1~kG rms on <$B$>) between our 
4 observing seasons, and detect no rotational modulation of <$B$> on measurements from individual spectra (which is unsurprising given the nearly pole-on viewing configuration of TW~Hya). } 

\section{Zeeman-Doppler Imaging of TW~Hya}
\label{sec:zdi}

In this section, we analyse the Stokes $I$ and $V$ LSD signatures of TW~Hya from each season using Zeeman-Doppler Imaging (ZDI), in order to simultaneously reconstruct 
the topology of the large-scale field and the associated brightness map, as well as their temporal evolution over the four seasons of our observations.  We achieve this 
through an iterative process that progressively adds information at the surface of the star, starting from a small magnetic seed and a featureless brightness map and 
exploring the parameter space with a variant of the conjugate gradient technique that aims at efficiently minimizing the discrepancy between the observed and synthetic 
Stokes $I$ and $V$ LSD profiles \citep[e.g.,][]{Skilling84, Brown91, Donati97c, Donati06b, Kochukhov16}.  Since the problem is ill posed, regularization is needed to 
ensure a unique solution.  ZDI uses the principles of maximum entropy image reconstruction, which aims at reaching a given agreement with the data, usually 
$\chisqr\simeq1$, while minimizing information in the derived maps to ensure that reconstructed features are mandatory to reproduce the data.  

In practice, we describe the surface of the star as a grid of 5000 cells and compute synthetic Stokes $I$ and $V$ profiles at each observation epoch by summing up the 
spectral contributions of all grid cells, taking into account the main geometrical parameters such as the cell coordinates, $i\simeq10$\degr, $\vsini=3$~\kms,  
and the linear limb darkening coefficient (set to 0.3).  We also assume that the surface of TW~Hya rotates as a solid body over each season, consistent with the low 
level of differential rotation observed on TW~Hya (see Secs.~\ref{sec:bl} and \ref{sec:rvs}).  Local Stokes $I$ and $V$ contributions from each cell are derived using 
Unno-Rachkovsky's analytical solution of the polarized radiative transfer equation in a plane-parallel Milne Eddington atmosphere \citep{Landi04}, assuming a Land\'e 
factor and average wavelength of 1.2 and 1750~nm for the LSD profiles, and a Doppler width $\vD=3$~\kms\ (including thermal, micro and macrotubulent broadening) for 
the local profile \citep[as for CI~Tau, see, e.g.,][]{Donati24}. 

The relative brightness at the surface of the star is described as a series of independent pixels, whereas magnetic field is expressed as a spherical harmonics (SH) 
expansion, using the formalism of \citet{Donati06b} in which the poloidal and toroidal components of the vector field depend on 3 sets of complex SH coefficients, 
$\alpha_{\ell,m}$ and $\beta_{\ell,m}$ for the poloidal component, and $\gamma_{\ell,m}$ for the toroidal component\footnote{We use here the modified 
expressions for the field components, where $\beta_{\ell,m}$ is replaced with $\alpha_{\ell,m}+\beta_{\ell,m}$ in the equations of the meridional and azimuthal field 
components \citep[see, e.g.,][]{Lehmann22, Finociety22, Donati23}.}, where $\ell$ and $m$ note the degree and order of the corresponding SH term in the expansion.  
Given the low \vsini\ of TW~Hya, we can safely limit the expansion to terms up to $\ell=5$.  As for other cTTSs magnetically imaged with ZDI to date, we favour 
large-scale field configurations that are mostly antisymmetric 
with respect to the centre of the star, in which accretion funnels linking the inner disc to the star are anchored at high latitudes, which is achieved in practice by 
penalizing even SH modes with respect to odd ones in the entropy function \citep[as in, e.g.,][]{Donati11}.  

Finally, we assume that only a fraction $f_V$ of each grid cell (called filling factor of the large-scale field, equal for all cells) contributes to Stokes $V$ profiles, 
with a magnetic flux over the cell equal to $B_V$ (i.e., a magnetic field within the magnetic portion of the cells equal to $B_V/f_V$).  Similarly, we assume that a 
fraction $f_I$ of each grid cell (called filling factor of the small-scale field, again equal for all cells) hosts small-scale fields of strength $B_V/f_V$ (i.e., with 
a small-scale magnetic flux over the cell equal to $B_I = B_V f_I/f_V$).  This simple model implies in particular that the small-scale field locally scales up with the 
large-scale field (with a scaling factor of $f_I/f_V$), which ensures at least that the resulting Zeeman broadening from small-scale fields is consistent with the 
reconstructed large-scale field.  As for the cTTS CI~Tau \citep{Donati24}, we set $f_I\simeq0.8$ \citep[consistent with our results, see Sec.~\ref{sec:bl}, and 
with those of][]{Yang05} and $f_V\simeq0.4$,  which yields a satisfactory fit to the observed Stokes $I$ and $V$ profiles of TW~Hya, and 
reproduces in particular the conspicuous triangular shape of the magnetically broadened Stokes $I$ LSD profiles.  

\begin{figure*}
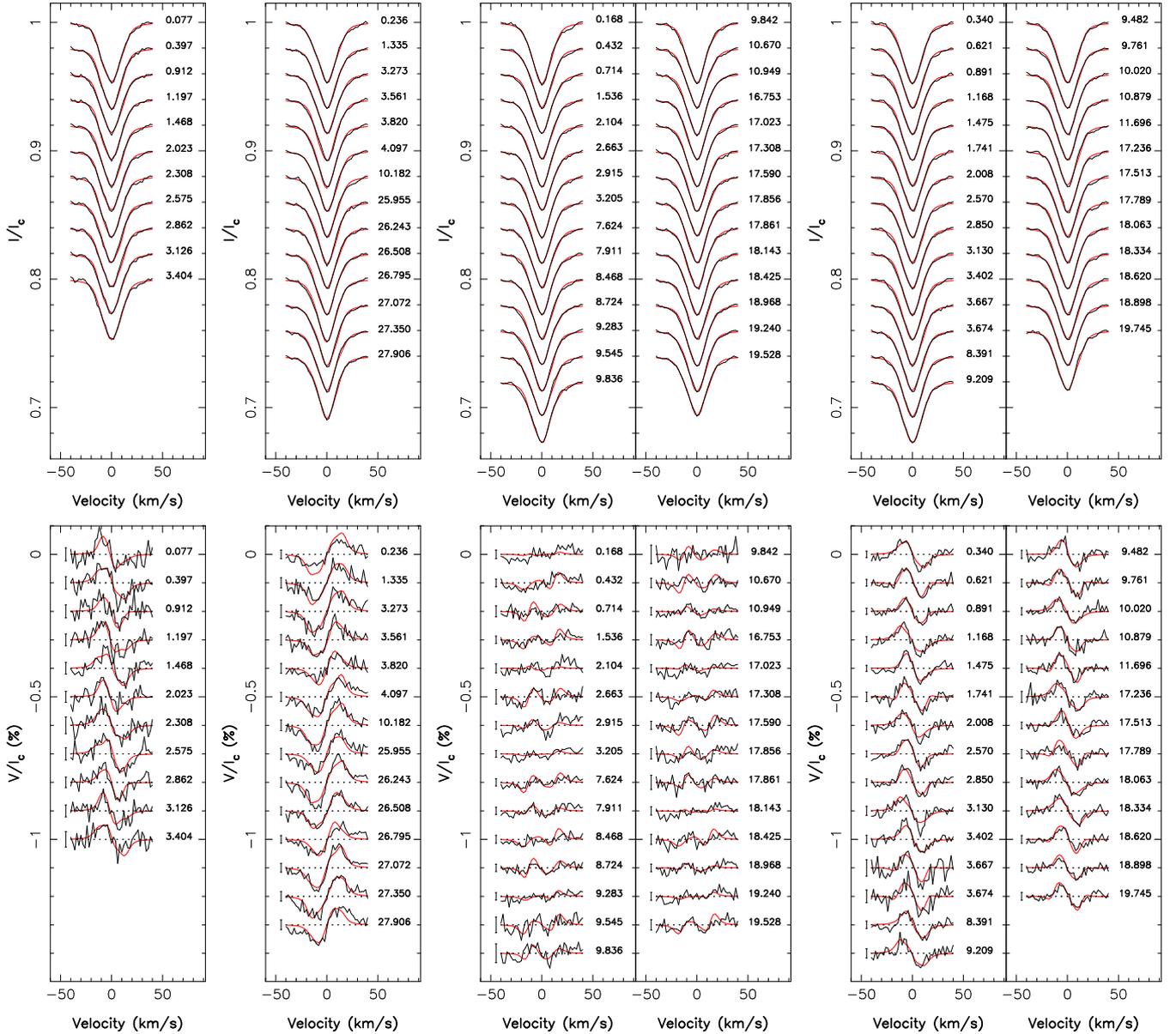

\centerline{\includegraphics[scale=0.45,angle=-90]{fig/twhya-fiti19.ps}\hspace{2mm}\includegraphics[scale=0.45,angle=-90]{fig/twhya-fiti20.ps}\hspace{2mm}\includegraphics[scale=0.45,angle=-90]{fig/twhya-fiti21.ps}\hspace{2mm}\includegraphics[scale=0.45,angle=-90]{fig/twhya-fiti22.ps}\vspace{2mm}} 
\centerline{\includegraphics[scale=0.45,angle=-90]{fig/twhya-fitv19.ps}\hspace{2mm}\includegraphics[scale=0.45,angle=-90]{fig/twhya-fitv20.ps}\hspace{2mm}\includegraphics[scale=0.45,angle=-90]{fig/twhya-fitv21.ps}\hspace{2mm}\includegraphics[scale=0.45,angle=-90]{fig/twhya-fitv22.ps}} 
\caption[]{Observed (thick black line) and modelled (thin red line) LSD Stokes $I$ (top row) and $V$ (bottom row) profiles of TW~Hya for seasons 2019 (first colum), 
2020 (second) column, 2021 (third colum) and 2022 (fourth column). {\emr Observed profiles were derived by applying LSD to our SPIRou spectra, using the atomic line mask 
outlined in Sec.~\ref{sec:obs}.}   
Rotation cycles (counting from 0, 82, 188 and 294 for the 2019, 2020, 2021 and 2022 seasons 
respectively, see Table~\ref{tab:log}) are indicated to the right of all LSD profiles, while $\pm$1$\sigma$ error bars are added to the left of Stokes $V$ signatures. } 
\label{fig:fit}
\end{figure*}

\begin{figure*}
\centerline{\large\bf 2019\raisebox{0.3\totalheight}{\includegraphics[scale=0.6,angle=-90]{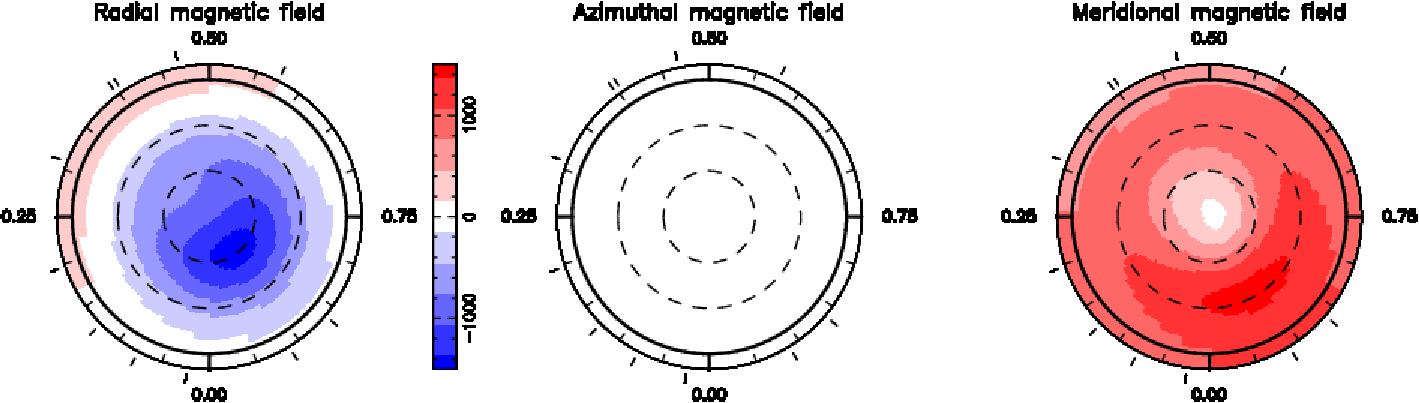}}\vspace{2mm}}
\centerline{\large\bf 2020\raisebox{0.3\totalheight}{\includegraphics[scale=0.6,angle=-90]{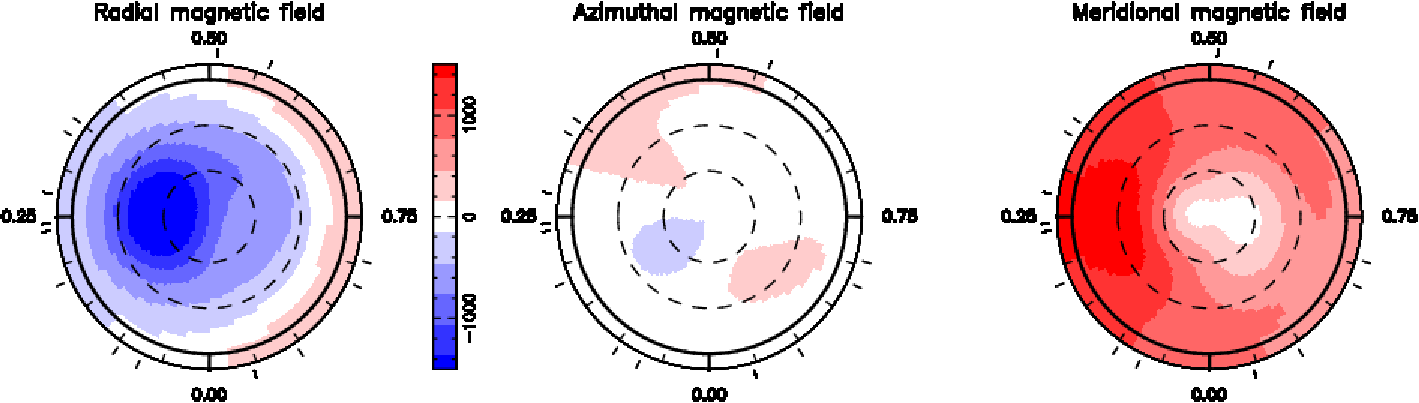}}\vspace{2mm}}
\centerline{\large\bf 2021\raisebox{0.3\totalheight}{\includegraphics[scale=0.6,angle=-90]{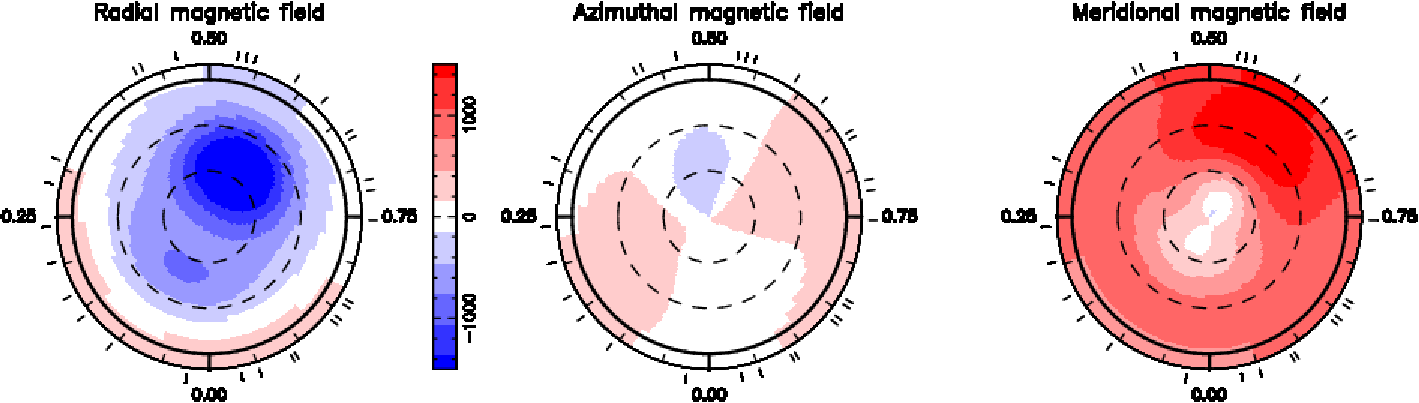}}\vspace{2mm}}
\centerline{\large\bf 2022\raisebox{0.3\totalheight}{\includegraphics[scale=0.6,angle=-90]{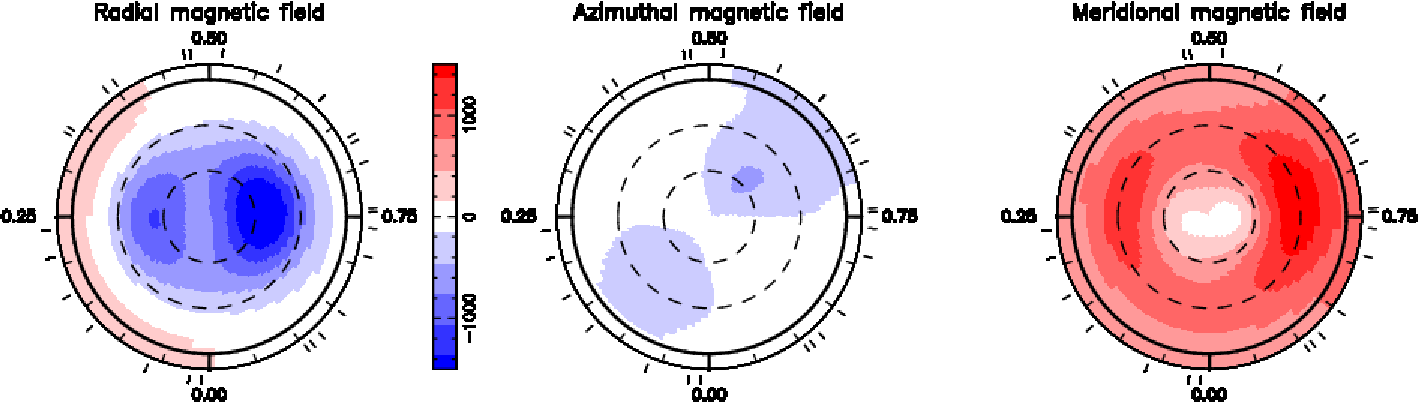}}}
\caption[]{Reconstructed maps of the large-scale field of TW~Hya (left, middle and right columns for the radial, azimuthal and meridional components in spherical 
coordinates, in G), for seasons 2019, 2020, 2021 and 2022 (top to bottom rows respectively), derived with ZDI from the Stokes $I$ and $V$ LSD profiles of Fig.~\ref{fig:fit}.  
The maps are shown in a flattened polar projection down to latitude $-10$\degr, with the north pole at the centre and the equator depicted as a bold line.  Outer ticks 
indicate phases of observations.  Positive radial, azimuthal and meridional fields respectively point outwards, counterclockwise and polewards. } 
\label{fig:map}
\end{figure*}

\begin{table} 
\caption[]{Properties of the large-scale and small-scale magnetic field of TW~Hya for our 4 observing seasons.  
We list the average reconstructed large-scale field strength <$B_V$> (column 2), the maximum small-scale field strength $B_I$ (column 3), the polar strength of the dipole 
component \Bd\ (column 4), the tilt / phase of the dipole field to the rotation axis (column 5) and the amount of magnetic energy reconstructed in the poloidal component of 
the field and in the axisymmetric modes of this component (column 6).  Error bars are typically equal to 5--10\% for field strengths and percentages, and 5--10\degr\ for 
field inclinations. } 
\begin{tabular}{cccccc}
\hline
Season & <$B_V$> & max $B_I$ & \Bd     & tilt / phase    & poloidal / axisym \\ 
       &  (G)    &   (kG)    &  (G)    & (\degr) & (\%)                      \\ 
\hline
2019   & 1050 & 2.9 & 1000 & 18 / 0.87 & 98 / 97 \\ 
2020   & 1130 & 3.9 & 1190 & 23 / 0.26 & 95 / 91 \\  
2021   & 1140 & 4.1 & 1130 & 20 / 0.57 & 96 / 91 \\  
2022   & 1030 & 3.5 &  990 & 17 / 0.74 & 98 / 93 \\
\hline 
\end{tabular}
\label{tab:mag}
\end{table}

Fitting our LSD Stokes $I$ and $V$ profiles with ZDI (down to $\chisqr\simeq1$, see Fig.~\ref{fig:fit}), we obtain reconstructed maps of the large-scale field, 
shown in Fig.~\ref{fig:map}.  Although brightness maps (not shown) were reconstructed at the same time as magnetic 
maps (with ZDI simultaneously fitting Stokes $I$ and $V$ profiles), we find that no brightness feature at the surface of TW~Hya is large enough, or exhibits a big 
enough nIR contrast with respect to the surrounding photosphere, to generate clear Stokes $I$ profile distortions and the corresponding rotational modulation, and 
thereby to show up in the derived brightness maps.  
This is in contrast with the brightness images we had derived from optical data showing an obvious dark feature coinciding with the magnetic pole  
\citep{Donati11}, but similar to our findings on CI~Tau where Stokes $I$ profile distortions induced by surface brightness features in the nIR were also barely detectable 
among the dominant ones induced by magnetic fields (conversely to the optical domain where the opposite behaviour holds).  

We find that TW~Hya hosts a large-scale magnetic field of average strength $\simeq$1.1~kG over the star and reaches a maximum intensity of 1.5--2.0~kG, which translates 
into average and maximum small-scale fields of 2.2 and 3--4~kG respectively (taking into account the $f_I/f_V\simeq2$ ratio between both quantities), consistent with literature 
values \citep[e.g.,][]{Yang05,Sokal18,Lopez-Valdivia21}.  The large-scale field  we reconstruct is almost fully poloidal and axisymmetric, and mainly consists of a 1.0--1.2~kG dipole 
inclined at 20\degr\ with respect to the rotation axis.  The octupole component is significantly weaker, with a polar strength ranging from 0.2 to 0.5~kG in the different seasons, 
and adds up to the polar large-scale field values, generating at times local maxima aside the main one in the radial field map, like in the 2021 and 2022 seasons.  
The main properties of the reconstructed magnetic topologies, consistent with the preliminary conclusions of our PCA analysis (see Sec.~\ref{sec:bl}), are summarized in Table~\ref{tab:mag}.  

We can see in particular that the large-scale magnetic topology is not undergoing global polarity switches over our 4-season timespan, despite the longitudinal field changing 
sign from 2019 to 2020 and again from 2021 to 2022 (see Fig.~\ref{fig:gpb} and Table~\ref{tab:log}).  It shows that, in a nearly pole-on viewing configuration like that of TW~Hya, 
sign switches in the longitudinal field may also reflect changes in the relative contributions of the radial field near the pole and the meridional field at lower latitudes (see 
Fig.~\ref{fig:map}), with the first dominating over the second when the strength of the dipole field is largest (i.e., in 2020 and 2021, see Table~\ref{tab:mag}).  

We note that the magnetic maps we derive from our SPIRou data differ from those reconstructed from ESPaDOnS data collected a decade earlier \citep{Donati11}, most likely as 
a result of changes in the large-scale field topology and accretion pattern, that we know can occur on relatively short timescales on TW~Hya \citep{Herczeg23}.  In particular, 
the octupole component measured from our SPIRou data is much smaller (by a factor of 5--10) than that derived from previous ESPaDOnS observations, whereas the dipole 
component is about twice larger \citep{Donati11}.  We estimate that most of this evolution is real, although we cannot exclude that some of it relates to the difference in the 
data sets and in particular in the wavelength domains.  Future analyses simultaneously combining optical and nIR spectropolarimetric data, such as that recently carried out for CI~Tau 
\citep{Donati24} should allow one to address this point in a more extensive way.  Despite such changes, the dipole tilts we measure from our new maps, of order 20\degr\ (see Table~\ref{tab:mag}), 
are consistent with the off-centring of the mainly polar accretion region taking place at the surface of TW~Hya, as derived from previous studies including ours \citep{Donati11, 
Sicilia23}.

\section{Radial velocity modeling of TW~Hya}
\label{sec:rvs}

Using now the spectra of TW~Hya reduced with APERO \citep{Cook22}, we can derive precise RVs with LBL \citep{Artigau22}, listed in Table~\ref{tab:log}, with a median RV precision 
of 2.2~\ms\ that reflects the relatively sharp lines of this star, {\emr and in particular the magnetically insensitive (unbroadened) molecular features}.  We find that TW~Hya is 
{\emr RV stable} at a rms level of 32.5~\ms\ over our 4 observing seasons, a dispersion about 15$\times$ larger than the median error bar on individual RV points.  By carrying out 
a GPR fit to these RVs, we find that 80\% of these variations are caused by rotational modulation, with a period equal to $3.5647\pm0.0024$~d, slightly but definitely smaller than 
the period derived from our \Bl\ data (see Sec.~\ref{sec:bl}).  This difference argues again for the presence of small latitudinal differential rotation at the surface of TW~Hya (at 
a level of a few \mrpd), with the polar regions contributing most to the \Bl\ variations (see Fig.~\ref{fig:map}) rotating more slowly than lower latitudes generating most of the RV 
modulation.  The semi-amplitude of this RV modulation is $25.5^{+5.6}_{-4.6}$~\ms, whereas the additional white noise on RVs (presumably caused by accretion-induced intrinsic 
distortions of spectral lines) reaches $19.0\pm1.7$~\ms, i.e., 8.6$\times$ the median error bar of our RV measurements (see Table~\ref{tab:pla}).  Note that two of the GPR hyper-parameters 
($\theta_3$ and $\theta_4$), weakly constrained by the data, were fixed to their optimal value from a preliminary run with all GPR parameters free to vary, making no difference 
on the filtering of activity.  

\begin{table*}
\caption[]{MCMC results for the 3 cases (no planet, planet b and planet b') of our RV analysis of TW~Hya. For each case, we list the recovered GP 
and planet parameters with their error bars, as well as the priors used whenever relevant.  The last 4 rows give the \chisqr\ and the rms 
of the best fit to our RV data, as well as the associated marginal logarithmic likelihood, $\log \mathcal{L}_M$, and marginal logarithmic 
likelihood variation, $\Delta \log \mathcal{L}_M$, with respect to the model without planet.  Two GPR hyper-parameters, weakly constrained by the data, 
were fixed to their optimal value from a preliminary run with all GPR parameters free to vary.  }
\begin{tabular}{cccccccccc}
\hline
Parameter          & No planet            & b                      & b'                     &   Prior \\
\hline
$\theta_1$ (\ms)   & $25.5^{+5.6}_{-4.6}$ & $25.2^{+5.4}_{-4.5}$   & $25.7^{+5.6}_{-4.6}$   & mod Jeffreys ($\sigma_{\rm RV}$) \\
$\theta_2$ (d)     & $3.5647\pm0.0024$    & $3.5649\pm0.0024$      & $3.5649\pm0.0023$      & Gaussian (3.56, 0.2) \\
$\theta_3$ (d)     & 300                  & 300                    & 300                    &  \\
$\theta_4$         & 0.6                  & 0.6                    & 0.6                    &  \\
$\theta_5$ (\ms)   & $19.0\pm1.7$         & $17.1\pm1.6$           & $17.1\pm1.6$           & mod Jeffreys ($\sigma_{\rm RV}$) \\
\hline
$K_b$ (\ms)        &                      & $11.1^{+3.3}_{-2.6}$   & $10.5^{+3.3}_{-2.5}$   & mod Jeffreys ($\sigma_{\rm RV}$) \\
$P_b$ (d)          &                      & $8.339\pm0.008$        & $8.147\pm0.010$        & Gaussian (8.34 or 8.15, 0.2) \\
BJD$_b$ (2459000+) &                      & $202.78\pm0.33$        & $204.90\pm0.37$        & Gaussian (202.8 or 204.9, 2.0) \\
$M_b \sin i$ (\me) &                      & $30^{+9}_{-7}$         & $28^{+9}_{-6}$         & derived from $K_b$, $P_b$ and \mstar \\
\hline
\chisqr            & 65.0                 & 51.4                     &  52.6   & \\
rms (\ms)          & 17.6                 & 15.6                     &  15.8   & \\
$\log \mathcal{L}_M$& 188.8               & 197.5                    & 196.9   & \\
$\log {\rm BF} = \Delta \log \mathcal{L}_M$ & 0.0     & 8.7          &   8.1   & \\
\hline
\end{tabular}
\label{tab:pla}
\end{table*}

\begin{figure*}
\centerline{\includegraphics[scale=0.6,angle=-90]{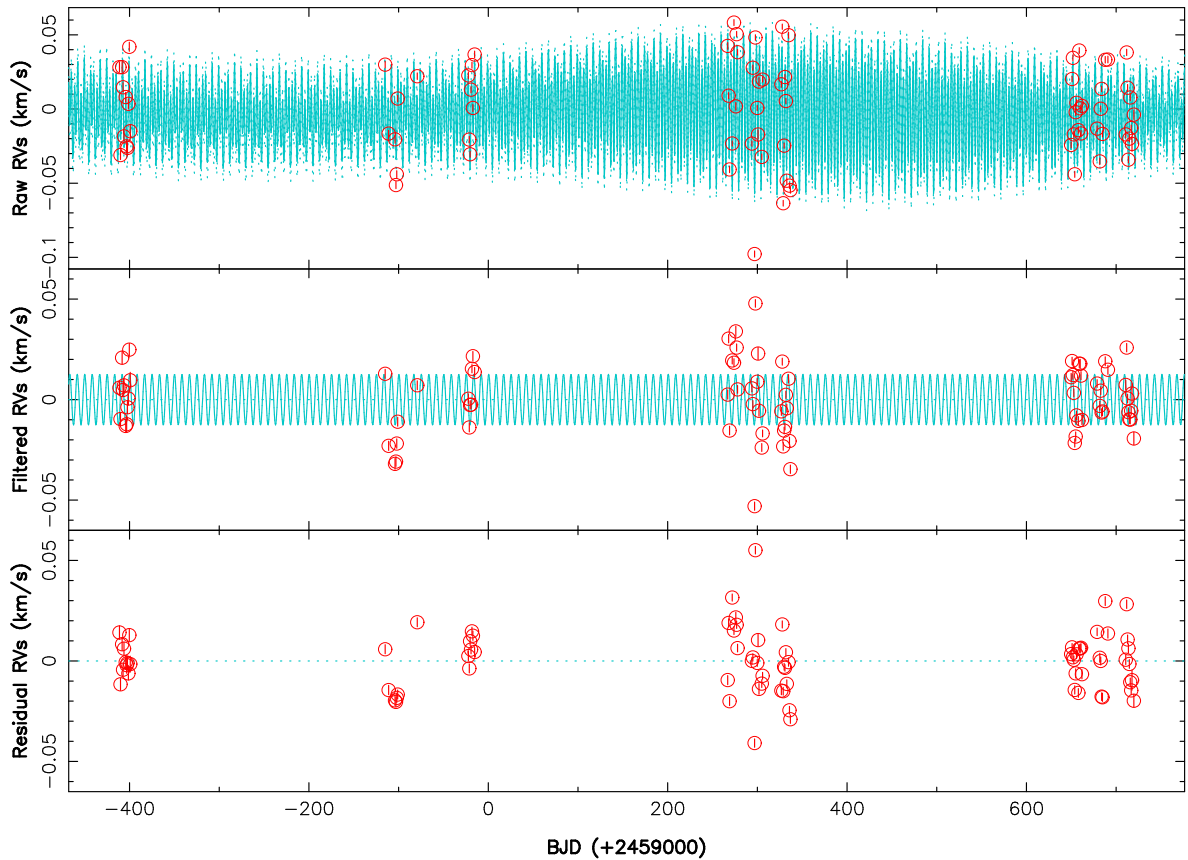}} 
\caption[]{Raw (top), filtered (middle) and residual (bottom) RVs of TW~Hya (red open circles).  The top plot shows the MCMC fit to the RV data, including 
a QP GPR modeling of the activity and the RV signature of a putative close-in planet of orbital period $8.339\pm0.008$~d (cyan), whereas the middle 
plot shows the planet RV signature alone once activity is filtered out.  The rms of the RV residuals is 15.6~\ms. } 
\label{fig:rvr}
\end{figure*}

\begin{figure}
\includegraphics[scale=0.48,angle=-90]{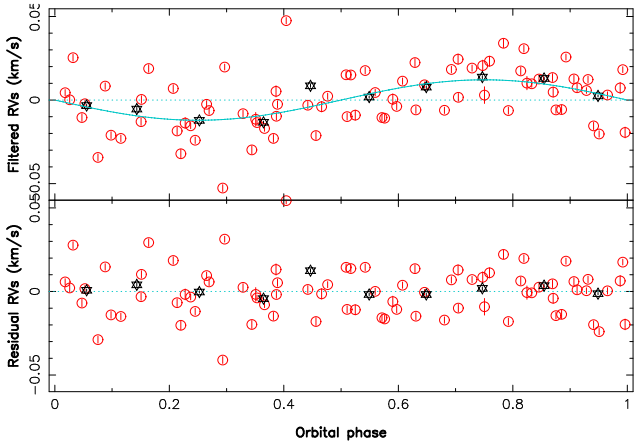}
\caption[]{Filtered (top plot) and residual (bottom plot) RVs of TW~Hya phase-folded on the 8.34-d period.  
The red open circles are the individual RV points with the respective error bars, whereas the black stars are average RVs over 0.1 phase bins.
As in Fig.~\ref{fig:rvr}, the dispersion of RV residuals is 15.6~\ms. }
\label{fig:rvf}
\end{figure}

Once activity is filtered out, we find residual power at a period of about 8.3~d in the RV periodogram.  We thus ran a new series of GPR fits to our RV points through a 
Bayesian Monte Carlo Markov chain experiment, including now the presence of a putative close-in planet in circular orbit around TW~Hya at a period of about 8.3~d described 
with 3 additional parameters.  We find that a relatively clear RV signal is present in the data at a 4.3$\sigma$ level, with a semi-amplitude of $11.1^{+3.3}_{-2.6}$~\ms\ 
and a period of $8.339\pm0.008$~d, corresponding to an orbital distance of $0.075\pm0.001$~au.  The RV residuals are now smaller (15.6~\ms\ instead of 17.1~\ms\ rms) than in the reference 
(no planet) case, and so is the additional white noise (17.1~\ms\ instead of 19.0~\ms).  The corresponding increase in marginal log likelihood $\log {\rm BF} = \Delta \log \mathcal{L}_M$ reaches 
8.7 (see Table~\ref{tab:pla}), suggesting that the detected RV signal is real \citep[i.e., $\log {\rm BF}>5$,][]{Jeffreys61}.  The corresponding fit to our RV points is shown in Fig.~~\ref{fig:rvr}, 
along with the filtered RVs, the fitted RV signal and the RV residuals.  The 1-yr alias of the reported RV signal, located at a period of $8.147\pm0.010$~d, is also a potential 
solution (which we refer to as case b' in Table~\ref{tab:pla}), albeit with a slightly lower confidence level ($\log {\rm BF}=8.1$).  
The corresponding periodograms are depicted in Fig.~\ref{fig:per}, along with the stacked periodogram illustrating how the main RV signal and its 1-yr alias get stronger 
and more dominant as more spectra are included in the analysis (see Fig.~\ref{fig:stp}).  The associated corner plot is shown in Fig.~\ref{fig:cor}.  We note that the 
period of this RV signal is slightly smaller than that of the photometric one that dominates the March 2021 and 2023 TESS light curves of TW~Hya (at about 9~d, see, e.g., 
Fig.~\ref{fig:tes}), though it is not clear whether both are related.  

If this RV signal is generated by a true planet, the minimum mass of this orbiting body would be $30^{+9}_{-7}$~\me\ (or $28^{+9}_{-6}$~\me\ in case b');  further assuming that 
the planet orbital plane coincides with the equatorial plane of the star yields a planet mass of $M_b = 0.55^{+0.17}_{-0.13}$~\mjup\ ($0.51^{+0.17}_{-0.12}$~\mjup\ in case b').  
The corresponding phase-folded RV curve is shown in Fig.~\ref{fig:rvf} for our main solution (with case b' yielding a very similar plot).  
Running again the same experiment assuming now a more general Keplerian orbit, we derive an eccentricity consistent with zero (error bar 0.04), and no obvious improvement in 
$\log {\rm BF}$ with respect to the circular orbit case.  

Due to the presence of a disc around TW~Hya, one may wonder whether the RV signal we detect truly comes from the reflex motion of the host star under the gravitational pull of a 
planet.  One could for instance suspect the disc itself to contribute to the spectral lines of TW~Hya, in particular the molecular lines that dominate the spectrum, and generate a 
modulated RV signal that would rather reflect, e.g., a non-axisymmetric structure within the disc rather than a genuine planet.  If this were the case, one would expect the molecular 
lines to be more affected than atomic lines, or lines in the $J$ and $H$ bands to be less impacted than those in the $K$ band, as a result of the different temperatures of the stellar 
photosphere and disc material.  We thus also analysed RVs obtained from Gaussian fits to Stokes $I$ LSD profiles of atomic lines only, that presumably come from the stellar photosphere 
with no contribution from the disc.  If the detected RV signal were not present in atomic lines \citep[as for CI~Tau, e.g.,][]{Donati24}, this would argue for it being induced by 
the disc.  However, atomic lines are 3$\times$ broader (as a consequence of magnetic fields) than molecular lines in TW~Hya, and telluric residuals affect LSD 
profiles more than LBL RV measurements.  As a result, the RV precision we obtain from atomic lines is significantly worse than in our main analysis, with an excess white noise from the GPR fit 
reaching 60~\ms\ (instead of 17~\ms\ when using LBL RVs from all spectral lines, see Table~\ref{tab:pla}).  Using LSD profiles from CO lines yields better results than from atomic lines 
(with an excess noise reduced by a factor of 2, down to 30~\ms), but still not good enough to unambiguously detect the RV signal detected from LBL measurements.  We also looked 
at the LBL RV measurements using lines from the $J$, $H$ and $K$ bands only, and again find that the excess noise (equal to 50, 32 and 25~\ms\ for the $J$, $H$ and $K$ bands respectively) is 
still too large to enable a firm detection of the RV signal in each band (with respective error bars of 10, 6 and 5~\ms\ on the semi-amplitude) and thus to look for potential differences 
between them.  It is therefore not possible at this stage to either confirm nor refute the planetary origin of the RV signal we report here.

\section{Emission lines of TW~Hya}
\label{sec:eml}

\begin{figure*}
\centerline{\hspace{-2mm}\includegraphics[scale=0.3,angle=-90]{fig/twhya-hei.ps}\hspace{19mm}\includegraphics[scale=0.3,angle=-90]{fig/twhya-pab.ps}\vspace{2mm}}
\centerline{\includegraphics[scale=0.55,angle=-90]{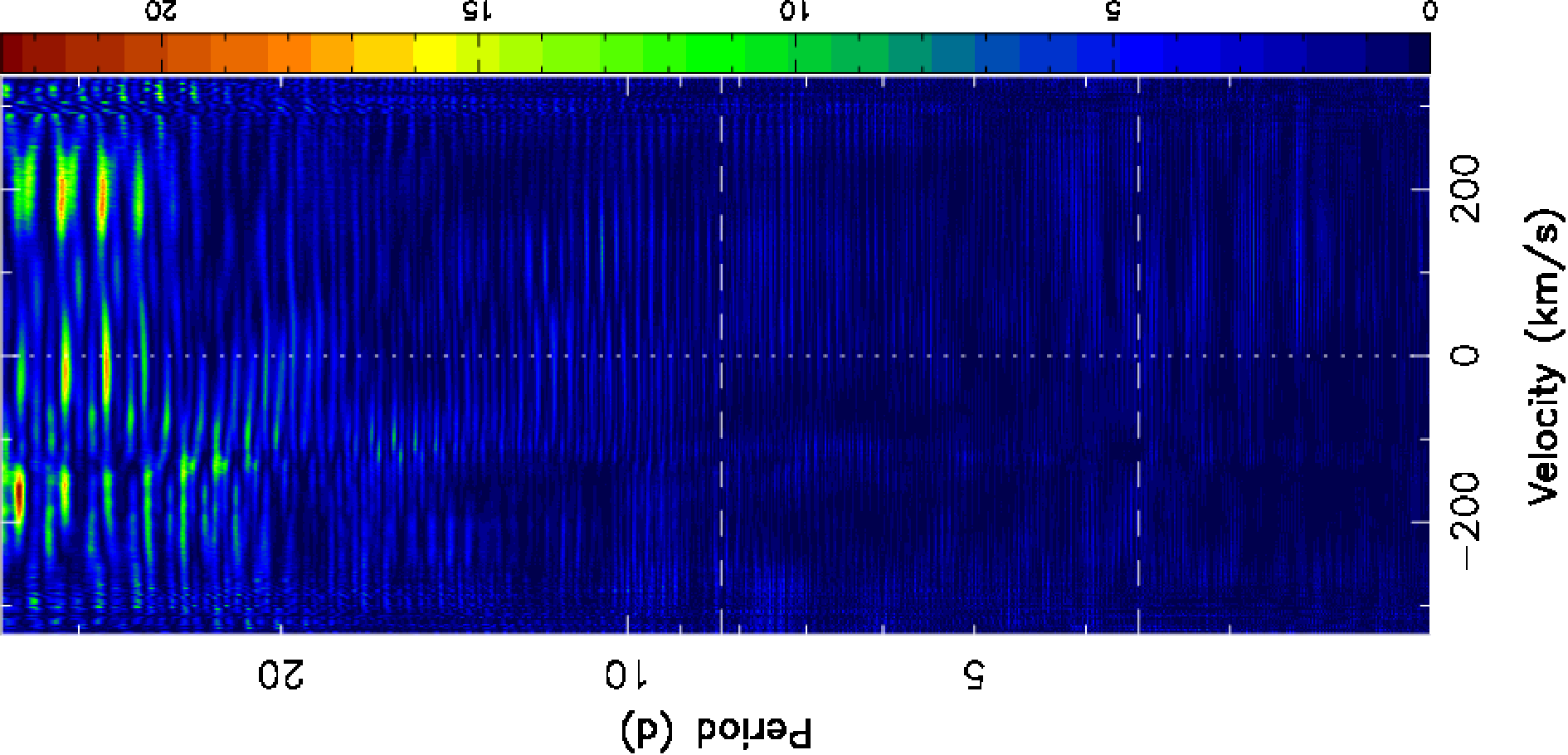}\hspace{3mm}\includegraphics[scale=0.55,angle=-90]{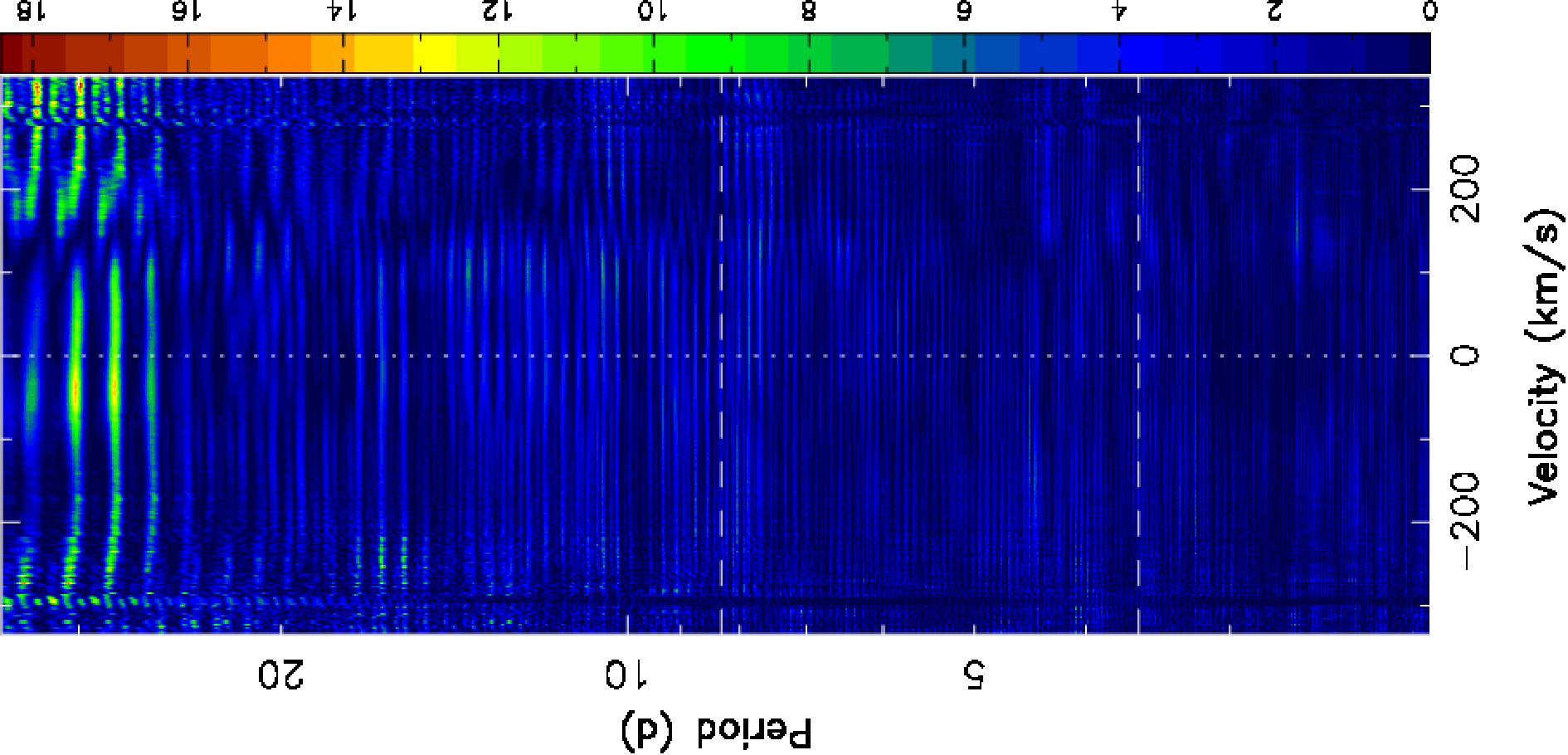}}
\caption[]{Stacked Stokes $I$ profiles and 2D periodograms of the 1083.3-nm \hei\ triplet (left panels) and of the 1282.16-nm \pab\ line (right) in the stellar 
rest frame, for our complete data set.  {\emr The color scale depicts the logarithmic power of the periodogram.}
Both lines are strongly variable with time, but with no obvious periodicity showing up, in particular at \Prot\ and $P_b$ (dashed horizontal 
lines), apart from peaks close to the synodic period of the Moon (29.5~d) and its 1-yr aliases (also present in the window function, see Fig.~\ref{fig:per}).  
} 
\label{fig:eml}
\end{figure*}

In this penultimate section, we discuss the main emission lines present in the nIR spectra of TW~Hya, and in particular the 1083.3-nm \hei\ triplet, as well 
as the 1282.16-nm \pab\ and 2166.12-nm \brg\ lines, known to probe accretion flows as well as outflows, in particular for the \hei\ triplet with its 
conspicuous P~Cygni profile featuring a broad and strong blue-shifted absorption component.  The stacked profiles and the associated 2D periodograms over 
the full data set are shown in Fig.~\ref{fig:eml} for \hei\ and \pab, whereas those of \brg\ are depicted in Fig.~\ref{fig:eml2}.  

We note that the \hei\ blue-shifted absorption, whose shape suggests it is formed within the stellar wind rather than from a disc wind \citep{Kwan07}, is 
strongly variable with time, sometimes extending blue-wards as far as $-300$~\kms\ but only down to $-150$~\kms\ at other epochs.  Its median equivalent width (EW) 
is about 100~\kms\ (0.36~nm, with no scaling from veiling).  With a median EW of about 150~\kms\ (0.54~nm), the emission component is also quite variable, sometimes 
dominating the whole profile and at other times almost non-existent.  About half the spectra show red-shifted absorption at velocities of 200~\kms, likely tracing 
accreted material from the disc falling onto the polar regions of TW~Hya.  

Neither the blue-shifted nor the red-shifted absorption components are modulated with rotation, even in individual seasons.  This may sound surprising at first glance, 
at least for the red-shifted absorption given the conclusion of Sec.~\ref{sec:zdi} that accretion occurs mostly towards the pole on TW~Hya in a more or less geometrically stable 
fashion.   However, TW~Hya being viewed almost pole-on, accreted material only comes in front of the stellar disc once close to the surface of the star where it ends up 
being visible all the time, thereby rendering rotational modulation much smaller than intrinsic variability.  This is likely the same for the blue-shifted component,  
especially if formed within a stellar wind;  alternatively, it may suggest that this component traces a wind from the inner disc (rather than from the star) for 
which no rotational modulation is expected, and no more than marginal evidence for longer periods is observed (apart from those attributable to the window function at the synodic 
period of the Moon, i.e., 29.5~d, and its 1-yr aliases, see Fig.~\ref{fig:per}).  
Such incoherent variability of emission lines is similar to what is seen in photometric light curves \citep[e.g.,][]{Rucinski08,Siwak14,Siwak18,Sicilia23}, 
including the 2019 and 2021 TESS light curves collected at the same time of our SPIRou observations (see Fig.~\ref{fig:tes} for their stacked periodograms).  

\begin{figure*}
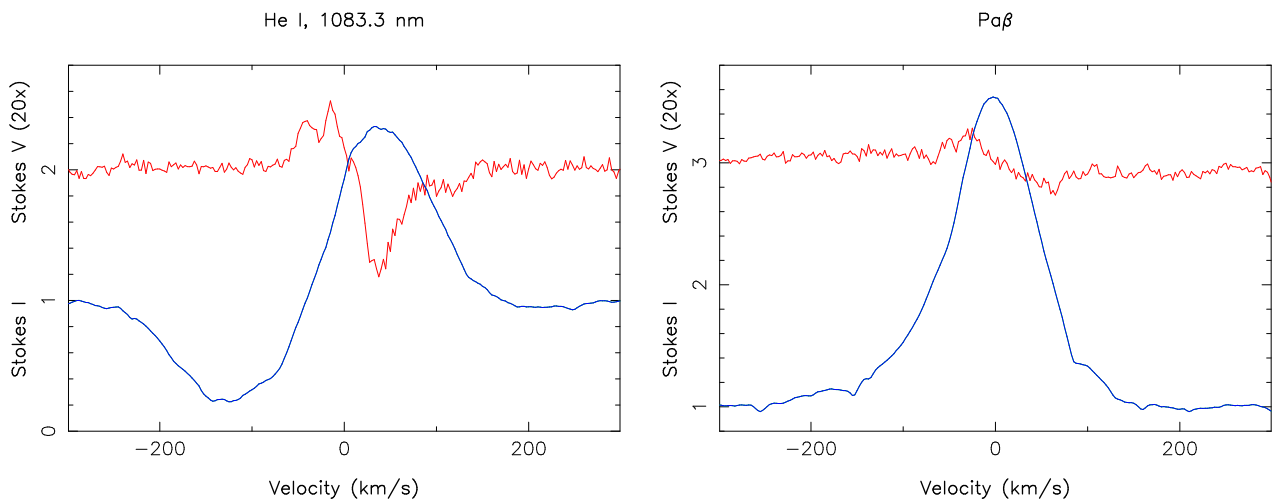

\centerline{\includegraphics[scale=0.35,angle=-90]{fig/twhya-heiv.ps}\hspace{5mm}\includegraphics[scale=0.35,angle=-90]{fig/twhya-pabv.ps}}
\caption[]{Weighted-average Stokes $I$ (blue) and $V$ (red) LSD profiles of the 1083.3-nm \hei\ (left) and \pab\ (right) lines of TW~Hya over our full data set. 
Zeeman signatures are clearly detected in both cases in conjunction with the emission peak.  The Stokes $V$ LSD profiles are both expanded by a factor of 20$\times$ 
and shifted upwards (by 2 and 3 respectively) for graphics purposes.  } 
\label{fig:emv}
\end{figure*}

We also note that a clear Stokes $V$ Zeeman signature is visible in the weighted average of all \hei\ profiles, as well as in those of each individual season.  
These signatures centred in the stellar rest frame and falling in conjunction with the emission peak (see Fig.~\ref{fig:emv}), demonstrate that at least part of 
the \hei\ emission comes from the footpoints of accretion funnels, i.e., close to where the large-scale field is strongest, and indicate the presence of an axisymmetric magnetic 
field component of negative polarity that is visible at all times.  This agrees well with our reconstructed ZDI maps that indeed show a negative radial field 
region close to the pole (see Fig.~\ref{fig:map}) and thereby always visible to the observer given the viewing angle of TW~Hya.  Assuming that the longitudinal 
field over the accretion region is similar to that previously probed by the 588~nm \hei\ $D_3$ line \citep[i.e., 2.5--3.0~kG, see][]{Donati11}, it implies 
that about 20\% of the emission flux in the 1083.3-nm \hei\ triplet, i.e., a component of EW $\simeq$30~\kms\ (0.11~nm), is coming from the post-shock accretion 
region within the chromosphere of TW~Hya.  

The \pab\ line of TW~Hya shows a simpler profile, with a main emission peak that is slightly blue-shifted (by --2.7~\kms\ in average) and features an extended 
blue wing.  The EW of \pab, measured through a simple Gaussian fit without any scaling for veiling, are listed in Table~\ref{tab:log} 
and vary from 120 to 616~\kms\ (0.51 to 2.63~nm), with a median of 300~\kms\ (1.28~nm).  {\emr When scaled up for veiling, these EWs translate into logarithmic 
luminosities (relative to \lsun) in \pab\ of $-3.79\pm0.18$ (with the error bar corresponding to temporal variability), and thus into logarithmic accretion 
luminosities (again relative to \lsun) of $-1.25\pm0.19$ \citep[following the calibrated relations of][]{Alcala17}.  This implies an average logarithmic mass 
accretion rate of $-8.48\pm0.19$ (in units of \mspy), with temporal variations in the range $-8.88$ to $-8.13$ (i.e., by a factor of 5.6 peak to 
peak).}  

The conspicuous absorption component that shows up in the red wing at a 
velocity of 84~\kms\ is caused by a photospheric line from Ti (located at 1282.52~nm).  The red-shifted absorption visible in the \hei\ line (tracing accreting 
material from the disc about to reach the stellar surface) is apparently also present in \pab, though much shallower.  The 2D periodogram of \pab\ over our 
full data set shows no clear feature apart from those attributable to the window function (see Fig.~\ref{fig:per}) and already present in the \hei\ periodogram.  
By running GPR through the EWs of \pab, we further confirm that no clear period emerges from the noise, dominated by accretion-induced 
intrinsic variability.  As for \hei, we speculate that the non-detection of rotational modulation is due to the viewing angle under which TW~Hya is seen from 
the Earth.  We also detect a clear Zeeman signature in conjunction with \pab\ (see Fig.~\ref{fig:emv}), albeit with a lower amplitude than that in \hei, and 
again probing the presence of an axisymmetric magnetic field component of negative polarity at the surface of TW~Hya.  Assuming this axisymmetric magnetic 
component is the same as that detected in \hei, it implies that the corresponding post-shock region in the chromosphere of TW~Hya contributes to the emission 
of \pab\ at an average EW of $\simeq$20~\kms\ (0.09~nm), i.e., about 7\% that of the whole \pab\ emission.  

Similar results are derived from \brg\ (see Fig.~\ref{fig:eml2}), with a comparable overall blue-shift (of --2.3~\kms) and EWs (listed in Table~\ref{tab:log}) 
that vary from 19 to 127~\kms\ (0.14 to 0.92~nm) with a median of 57~\kms\ (0.41~nm).  
{\emr These EWs translate into logarithmic luminosities in \brg\ of $-4.83\pm0.21$ and into logarithmic accretion luminosities of $-1.73\pm0.24$ (both relative to \lsun), 
implying an average logarithmic mass accretion rate of $-8.96\pm0.24$ (in units of \mspy, with temporal variations in the range $-9.51$ to $-8.53$, 
i.e., by a factor of 9.5 peak to peak).} A Zeeman signature is again detected in \brg\ with the same characteristics 
as that of \pab;  assuming once more that it probes the same axisymmetric magnetic component (of negative polarity), we can infer that the post-shock region at 
the footpoint of accretion funnels contributes to the emission of \brg\ at an average EW of 7~\kms\ (0.05~nm), i.e., about 12\% that of the whole \brg\ emission.  

{\emr The average logarithmic mass-accretion rate that we derive for TW~Hya from our 2019 to 2022 SPIRou observations (using both \pab\ and \brg) is thus equal 
to $-8.72\pm0.22$ (in units of \mspy, with temporal variations in the range $-9.19$ to $-8.34$) in good agreement with the results of \citet{Herczeg23} derived 
from 2.5 decades of irregular monitoring at optical wavelengths. } 

We finally note that no power is detected in either lines at the period of the RV signal reported in Sec.~\ref{sec:rvs}, which is what we expect if the RV signal 
is not attributable to activity.  However, as no power is detected at \Prot\ either, where one usually expects activity to show up, the non-detection at $P_b$ 
does not provide definitive evidence that this period is unrelated to activity.  

\section{Summary and discussion}
\label{sec:dis}

We monitored the cTTS TW~Hya with the SPIRou high-resolution nIR spectropolarimeter / velocimeter at CFHT over 4 consecutive seasons (from 2019 to 2022), 
in the framework of the SLS large program and of a PI program.  We obtained a total of 82 usable Stokes $I$ and $V$ spectra of TW~Hya 
on which the LSD and LBL methods were applied to derive Zeeman signatures and precise RVs for each of our observing nights.  

The longitudinal field \Bl\ measured from Stokes $I$ and $V$ LSD profiles evolved with time, and even switched sign between 2019 and 2020, and again between 
2021 and 2022.  Rotational modulation of \Bl\ is also detected, yielding a period equal to $\Prot=3.606\pm0.015$~d, slightly but significantly larger than 
that derived from optical RVs collected in 2008 \citep[i.e., $3.5683\pm0.0002$~d,][]{Huelamo08}.  We also detect rotational modulation of RVs in our nIR data, 
with a period of $3.5649\pm0.0024$~d, consistent at 1.4$\sigma$ with the estimate from optical RVs.  It demonstrates that latitudinal differential rotation 
is present at the surface of TW~Hya, with the polar regions (mostly probed by \Bl) rotating more slowly than lower latitudes (to which RVs are mostly sensitive), 
but at a level of only a few \mrpd\ between the equator and pole, i.e., consistent with previous results on TTSs similar to TW~Hya \citep[e.g.,][]{Finociety23}.  
By modeling the Zeeman broadening of atomic and molecular lines of TW~Hya, we also measured a small-scale field of $3.60\pm0.04$~kG (with $7\pm2$\% of the 
visible stellar surface free of such field) and only small season-to-season variations, in rough agreement with previous literature 
estimates \citep{Yang05,Sokal18,Lavail19,Lopez-Valdivia21}.  

By carrying out a PCA analysis of our Stokes $V$ profiles \citep[as advocated by][]{Lehmann22,Lehmann24}, we find that the large-scale field of TW~Hya is 
mostly poloidal and axisymmetric at all epochs.  This conclusion is confirmed with a thorough modeling with ZDI, thanks to which 
we reconstructed the large-scale magnetic field of TW~Hya, as well as the photospheric brightness at nIR wavelengths, for each observing season, from 
a simultaneous fit to the corresponding set of Stokes $I$ and $V$ LSD profiles.  We find that the large-scale field is fairly stable with time, despite the 
sign switches in \Bl, with a dominant dipole component evolving from 1.0~kG (in 2019 and 2022) to 1.2~kG (in 2020) and 1.1~kG (in 2021).  The sign switches 
that \Bl\ exhibits directly reflects this temporal evolution of the large-scale field, with the polar and lower latitude regions both contributing to \Bl\ 
through the radial and meridional field respectively, in the mostly pole-on viewing configuration of TW~Hya.  Besides, we find that the nIR brightness 
inhomogeneities at the surface of TW~Hya only feature a low contrast with respect to the quiet photosphere, generating, along with the large-scale field, a 
rotational modulation of LBL RVs (from the narrower molecular lines mainly) whose semi-amplitude is only $25\pm5$~\ms, i.e., much lower than that from TTSs with high levels of 
spot coverage \citep[e.g.,][]{Finociety23}.  We also find that the small-scale fields derived from the shape and width of LSD profiles of atomic lines of TW~Hya are 
consistent with typical values of the small-scale and large-scale filling factors of cTTSs, i.e., $f_I\simeq0.8$ and $f_V\simeq0.4$ \citep[e.g.,][]{Donati24}, yielding 
average and maximum values of 2.2 and 3--4~kG respectively for the small-scale field at the surface of the star, again consistent with previous measurements including ours.  

Comparing with our previous large-scale magnetic field maps of TW~Hya from optical spectropolarimetric data collected a decade ago \citep{Donati11}, we find a clear 
evolution, with the dipole component about twice stronger and the octupole component much weaker (by a factor of 5--10) than it used to be.  We believe that 
most of this evolution is real, but cannot exclude that some of it reflects the difference in wavelength domains between both data sets.  Similarly, we note 
that the semi-amplitude of the RV modulation, equal to $25\pm5$~\ms\ in our nIR data, is an order of magnitude smaller than that reported from optical RVs 
collected 1.5~decades ago \citep[][]{Huelamo08, Donati11}, which again argues for intrinsic variability of the magnetic activity at the surface of TW~Hya, 
the typical ratio between the nIR and optical RV jitter being usually much smaller than 10 \citep[e.g.,][]{Mahmud11, Crockett12}.  These results strongly 
argue in favour of collecting spectropolarimetric and velocimetric data in both optical and nIR domains at the same time so that one can simultaneously use 
information from both spectral ranges, as recently done in the case of the cTTS CI~Tau \citep{Donati24}.  This should be routinely possible in 
a few months once SPIRou and ESPaDOnS are merged into a single instrument (called VISION) allowing one to simultaneously observe the same star in both 
wavelength domains.  

Given the reported mass accretion rate at the surface of TW~Hya over the previous 2 decades \citep[with $\log\Mdot\simeq-8.65$~\mspy, ranging from --9.2 to --8.2,][{\emr and consistent 
with the one we derive from our SPIRou observations, see Sec.~\ref{sec:eml}}]{Herczeg23}, 
we find that the magnetic truncation radius \rmag\ \citep[as defined by][and up to which the large-scale field is able to disrupt the Keplerian disc]{Bessolaz08} is equal to 
$\rmag=4.5^{+2.0}_{-1.1}$~\rstar, i.e., $0.57^{+0.25}_{-0.15}$~\rcor, the error bar reflecting mostly the reported variation in mass accretion rate rather than that in the dipole 
component of the large-scale field.  Our result is consistent with the recent interferometric measurement of the magnetospheric radius of TW~Hya, equal to $4.50\pm0.26$~\rstar\ 
\citep{Gravity20} at the time of their observations.  The average value of $\rmag/\rcor\simeq0.57$ means in particular that the dipole field of TW~Hya is not strong enough to spin the 
star down given the average mass accretion rate \citep{Zanni13,Pantolmos20}, consistent with the rotation period of TW~Hya being shorter than that of most prototypical cTTSs \citep[e.g., 
CI~Tau,][]{Donati20b}.  
However, the dipole field is nonetheless sufficiently intense to ensure that the accretion pattern is geometrically stable, with magnetic funnels linking the inner accretion disc to the 
polar regions at the surface of the star \citep[e.g.,][]{Sicilia23}, rather than to lower stellar latitudes through chaotic accretion tongues \citep{Blinova16}, at least when the 
accretion rate is not too large.  We suspect that previous reports of equator-ward accretion at the surface of  TW~Hya \citep[e.g.,][]{Argiroffi17} correspond to epochs where the 
accretion rate was close to its maximum.  We note that no rotational modulation is detected in the 1083.3-nm \hei, \pab\ and \brg\ accretion / ejection proxies despite our conclusion 
that the accretion pattern is geometrically stable.  This likely reflects the specific viewing angle of TW~Hya, causing rotational modulation to be minimal and thus easily hidden behind 
accretion-induced intrinsic temporal variability.  
Besides, we report the detection of Zeeman signatures in the \hei, \pab\ and \brg\ lines of TW~Hya, suggesting that 7--20\% of the line fluxes come from the hot chromospheric region 
at the footpoints of accretion funnels.  

Last but not least, we report that RVs of TW~Hya are also modulated with a period of $8.339\pm0.008$~d (or its 1-yr alias $8.147\pm0.010$~d), with a semi-amplitude 
of $11.1^{+3.3}_{-2.6}$~\ms.  This modulation may reflect the presence of a planet in a circular orbit around TW~Hya at a distance of $0.075\pm0.001$~au, i.e., 
beyond both the magnetospheric and corotation radii, and that would be detected with a confidence level of 4.3$\sigma$ ($\log {\rm BF}=8.7$).  If the orbit of this 
candidate planet is coplanar with the rotation of the star, this would imply a planet mass of $0.55^{+0.17}_{-0.13}$~\mjup.  An alternative option is that the RV 
signal we detect is caused by a non-axisymmetric density structure in the inner disc of TW~Hya (possibly induced by a lower mass planet), generating a spectral 
contribution to some of the spectral lines (e.g., the molecular lines) and thereby inducing a small amplitude modulation in the measured RVs.  We note that the period 
of this RV signal is slightly smaller than the photometric one that dominates the 2021 and 2023 TESS light curves;  it is too early to speculate whether both are physically 
related (e.g., with a planet or disc structure regularly triggering enhanced accretion), or rather simply coincide by chance.  
Additional SPIRou observations are needed to further investigate the spectral properties of this RV 
signal and unambiguously diagnose its origin, before claiming the detection of a close-in massive planet orbiting TW~Hya.  If confirmed, this detection would 
demonstrate that planet formation and migration is actively going on within the protoplanetary disc of TW~Hya and likely participates in generating the reported disc 
structures as previously suspected \citep[e.g.,][]{Tsukagoshi19}.  {\emr In particular, the candidate inner planet we report here may possibly be at the origin of the 
innermost gap at 1~au \citep{Andrews16} in the disc of TW~Hya and / or contribute to the misalignment of the inner disc rings within 7~au \citep{Debes23}.  It is however 
unlikely to have caused the more distant multiple gaps \citep[e.g., at about 25 and 40~au,][]{vanBoekel17,Huang18} in the outer disc, that may probe the presence of 
additional embedded massive planets within the protoplanetary disc of TW~Hya. }

\section*{Acknowledgements}
We thank an anonymous referee for valuable comments on the manuscript.  
This project received funds from the European Research Council (ERC) under the H2020 research \& innovation program (grant agreements \#740651 NewWorlds, \#742095 SPIDI) 
and under the Horizon Europe research \& innovation program (\#101053020  Dust2Planets).  SHPA acknowledges funding from FAPEMIG, CNPq and CAPES.  
This work was also supported by the NKFIH excellence grant TKP2021-NKTA-64, and 
benefited from the SIMBAD CDS database at URL {\tt http://simbad.u-strasbg.fr/simbad} and the ADS system at URL {\tt https://ui.adsabs.harvard.edu}.
Our study is based on data obtained at the CFHT, operated by the CNRC (Canada), INSU/CNRS (France) and the University of Hawaii.
The authors wish to recognise and acknowledge the very significant cultural role and reverence that the summit of Maunakea has always had
within the indigenous Hawaiian community.  We are most fortunate to have the opportunity to conduct observations from this mountain.

\section*{Data availability}  Data used in this paper are publicly available from the Canadian Astronomy Data Center.

\bibliography{twhya-arxiv} 
\bibliographystyle{mnras}

\appendix
{\emr 
\section{Abbreviations}
\label{sec:appD}

In Table~\ref{tab:abb}, we recall, by alphabetical order, all abbreviations used in the paper.  

\begin{table}
\caption[]{Abbreviations used in the paper, in alphabetical order}
\begin{tabular}{ll}
\hline
BF & Bayes factor \\ 
CFHT & Canada-France-Hawaii Telescope \\
cTTS & classical T Tauri star \\ 
EW & equivalent width \\ 
GP & Gaussian process \\ 
GPR & Gaussian process regression \\ 
LBL & line-by-line method \\ 
LSD & Least-Squares Deconvolution \\ 
MCMC & Monte Carlo Markov chain \\ 
nIR & near infrared \\ 
PCA & Principal Component Analysis \\ 
QP & quasi periodic \\ 
RV & radial velocity \\ 
SH & spherical harmonics \\ 
SLS & SPIRou Legacy Survey \\ 
SNR & signal-to-noise ratio \\ 
TTS & T Tauri star \\ 
TWA & TW Hydra Association \\ 
ZDI & Zeeman-Doppler imaging \\ 
\hline
\end{tabular}
\label{tab:abb}
\end{table}
} 

\section{Observation log}
\label{sec:appA}

Table~\ref{tab:log} gives the full log and associated \Bl\ and RV measurements at each observing epoch from our SPIRou spectra.  

\begin{table*} 
\small
\caption[]{Observing log of our SPIRou observations of TW~Hya in seasons 2019, 2020, 2021 and 2022.  All exposures consist of 4 sub-exposures of equal length.
For each visit, we list the barycentric Julian date BJD, the UT date, the rotation cycle c and phase $\phi$ (computed as indicated in Sec.~\ref{sec:obs}), 
the total observing time t$_{\rm exp}$, the peak SNR in the spectrum (in the $H$ band) per 2.3~\kms\ pixel, the noise level in the LSD Stokes $V$ profile, 
the estimated \Bl\ with error bars, the nightly averaged RVs and corresponding error bars derived by APERO and LBL, and finally the EWs of the \pab\ and \brg\ 
emission lines with error bars (no scaling up from veiling).  }  
\begin{tabular}{cccccccccc} 
\hline
BJD        & UT date & c / $\phi$ & t$_{\rm exp}$ & SNR & $\sigma_V$            & \Bl\   &  RV    & EW \pab\ & EW \brg\ \\ 
(2459000+) &         &            &   (s)        & ($H$) & ($10^{-4} I_c$)       & (G)   & (\ms)  & (\kms)   & (\kms)   \\  
\hline
-411.2218803 & 15 Apr 2019 & 0 / 0.077 & 1136.7 & 168 & 3.03 & 34$\pm$26 & 28$\pm$3 & 303$\pm$15 & 65$\pm$15 \\
-410.0666365 & 16 Apr 2019 & 0 / 0.397 & 1136.7 & 191 & 2.62 & 77$\pm$25 & -31$\pm$3 & 539$\pm$15 & 127$\pm$15 \\
-408.2102656 & 18 Apr 2019 & 0 / 0.912 & 1136.7 & 164 & 3.03 & 26$\pm$26 & 28$\pm$3 & 365$\pm$15 & 82$\pm$15 \\
-407.1835982 & 19 Apr 2019 & 1 / 0.197 & 1136.7 & 182 & 2.68 & 65$\pm$23 & 15$\pm$3 & 345$\pm$15 & 73$\pm$15 \\
-406.2071730 & 20 Apr 2019 & 1 / 0.468 & 1136.7 & 180 & 2.74 & 4$\pm$24 & -18$\pm$3 & 344$\pm$15 & 70$\pm$15 \\
-404.2046516 & 22 Apr 2019 & 2 / 0.023 & 1136.7 & 185 & 2.65 & 4$\pm$22 & 8$\pm$3 & 220$\pm$15 & 40$\pm$15 \\
-403.1758960 & 23 Apr 2019 & 2 / 0.308 & 1136.7 & 144 & 3.79 & 69$\pm$34 & -25$\pm$4 & 278$\pm$15 & 60$\pm$15 \\
-402.2162162 & 24 Apr 2019 & 2 / 0.575 & 1136.7 & 177 & 2.81 & 44$\pm$24 & -26$\pm$3 & 328$\pm$15 & 70$\pm$15 \\
-401.1785924 & 25 Apr 2019 & 2 / 0.862 & 1136.7 & 198 & 2.48 & 32$\pm$21 & 4$\pm$3 & 372$\pm$15 & 73$\pm$15 \\
-400.2281200 & 26 Apr 2019 & 3 / 0.126 & 1136.7 & 184 & 2.64 & 50$\pm$24 & 42$\pm$3 & 506$\pm$15 & 96$\pm$15 \\
-399.2256117 & 27 Apr 2019 & 3 / 0.404 & 1136.7 & 143 & 3.62 & 37$\pm$31 & -15$\pm$4 & 197$\pm$15 & 36$\pm$15 \\
\hline
-114.9575691 & 05 Feb 2020 & 82 / 0.236 & 1582.4 & 286 & 1.56 & -144$\pm$14 & 30$\pm$2 & 588$\pm$15 & 115$\pm$15 \\
-110.9945004 & 09 Feb 2020 & 83 / 0.335 & 1582.4 & 257 & 1.77 & -157$\pm$15 & -17$\pm$2 & 552$\pm$15 & 105$\pm$15 \\
-104.0041815 & 16 Feb 2020 & 85 / 0.273 & 1582.4 & 213 & 2.24 & -182$\pm$19 & -20$\pm$3 & 599$\pm$15 & 124$\pm$15 \\
-102.9682410 & 17 Feb 2020 & 85 / 0.561 & 1582.4 & 252 & 1.80 & -123$\pm$16 & -51$\pm$2 & 586$\pm$15 & 120$\pm$15 \\
-102.0330761 & 18 Feb 2020 & 85 / 0.820 & 1582.4 & 267 & 1.67 & -102$\pm$15 & -44$\pm$2 & 570$\pm$15 & 110$\pm$15 \\
-101.0329883 & 19 Feb 2020 & 86 / 0.097 & 1582.4 & 285 & 1.56 & -169$\pm$14 & 7$\pm$2 & 616$\pm$15 & 122$\pm$15 \\
-79.0917046 & 12 Mar 2020 & 92 / 0.182 & 1582.4 & 287 & 1.93 & -189$\pm$16 & 22$\pm$2 & 489$\pm$15 & 96$\pm$15 \\
-22.2151786 & 08 May 2020 & 107 / 0.955 & 1582.4 & 260 & 1.93 & -175$\pm$17 & 23$\pm$2 & 496$\pm$15 & 99$\pm$15 \\
-21.1748547 & 09 May 2020 & 108 / 0.243 & 1582.4 & 275 & 1.68 & -195$\pm$15 & -21$\pm$2 & 401$\pm$15 & 80$\pm$15 \\
-20.2194204 & 10 May 2020 & 108 / 0.508 & 2005.9 & 333 & 1.38 & -129$\pm$12 & -30$\pm$2 & 387$\pm$15 & 76$\pm$15 \\
-19.1862407 & 11 May 2020 & 108 / 0.795 & 2005.9 & 335 & 1.41 & -133$\pm$12 & 13$\pm$2 & 301$\pm$15 & 54$\pm$15 \\
-18.1874428 & 12 May 2020 & 109 / 0.072 & 2005.9 & 315 & 1.49 & -163$\pm$13 & 30$\pm$2 & 504$\pm$15 & 107$\pm$15 \\
-17.1830447 & 13 May 2020 & 109 / 0.350 & 2005.9 & 305 & 1.65 & -184$\pm$15 & 1$\pm$2 & 528$\pm$15 & 106$\pm$15 \\
-15.1805270 & 15 May 2020 & 109 / 0.906 & 2005.9 & 281 & 1.69 & -182$\pm$15 & 37$\pm$2 & 326$\pm$15 & 57$\pm$15 \\
\hline
267.0324755 & 21 Feb 2021 & 188 / 0.168 & 2005.9 & 318 & 1.51 & -105$\pm$13 & 43$\pm$2 & 136$\pm$15 & 25$\pm$15 \\
267.9869082 & 22 Feb 2021 & 188 / 0.432 & 2005.9 & 346 & 1.34 & -112$\pm$11 & 9$\pm$2 & 135$\pm$15 & 24$\pm$15 \\
269.0011213 & 23 Feb 2021 & 188 / 0.714 & 2005.9 & 278 & 1.73 & -48$\pm$16 & -41$\pm$2 & 249$\pm$15 & 52$\pm$15 \\
271.9683085 & 26 Feb 2021 & 189 / 0.536 & 2005.9 & 379 & 1.23 & -75$\pm$11 & -23$\pm$2 & 258$\pm$15 & 50$\pm$15 \\
274.0167096 & 28 Feb 2021 & 190 / 0.104 & 2005.9 & 325 & 1.52 & -78$\pm$13 & 58$\pm$2 & 272$\pm$15 & 51$\pm$15 \\
276.0307067 & 02 Mar 2021 & 190 / 0.663 & 2005.9 & 246 & 2.49 & -49$\pm$21 & 2$\pm$3 & 176$\pm$15 & 29$\pm$15 \\
276.9398811 & 03 Mar 2021 & 190 / 0.915 & 2005.9 & 338 & 1.29 & -97$\pm$11 & 50$\pm$2 & 217$\pm$15 & 43$\pm$15 \\
277.9841904 & 04 Mar 2021 & 191 / 0.205 & 2005.9 & 392 & 1.15 & -74$\pm$10 & 38$\pm$2 & 139$\pm$15 & 21$\pm$15 \\
293.9200402 & 20 Mar 2021 & 195 / 0.624 & 2005.9 & 383 & 1.20 & -68$\pm$10 & -23$\pm$2 & 168$\pm$15 & 25$\pm$15 \\
294.9534459 & 21 Mar 2021 & 195 / 0.911 & 2005.9 & 362 & 1.25 & -68$\pm$11 & 28$\pm$2 & 123$\pm$15 & 19$\pm$15 \\
296.9619202 & 23 Mar 2021 & 196 / 0.468 & 2005.9 & 392 & 1.16 & -60$\pm$12 & -98$\pm$2 & 375$\pm$15 & 92$\pm$15 \\
297.8864493 & 24 Mar 2021 & 196 / 0.724 & 2005.9 & 401 & 1.13 & -35$\pm$11 & 48$\pm$2 & 264$\pm$15 & 60$\pm$15 \\
299.9019941 & 26 Mar 2021 & 197 / 0.283 & 2005.9 & 363 & 1.24 & -64$\pm$11 & 1$\pm$2 & 207$\pm$15 & 45$\pm$15 \\
300.8485035 & 27 Mar 2021 & 197 / 0.545 & 2005.9 & 188 & 2.63 & -81$\pm$23 & -17$\pm$3 & 162$\pm$15 & 33$\pm$15 \\
301.8981152 & 28 Mar 2021 & 197 / 0.836 & 2005.9 & 190 & 3.15 & -71$\pm$28 & 18$\pm$3 & 120$\pm$15 & 23$\pm$15 \\
301.9183606 & 28 Mar 2021 & 197 / 0.842 & 2005.9 & 187 & 3.00 & -48$\pm$27 & & 125$\pm$15 & 22$\pm$15 \\
304.9025306 & 31 Mar 2021 & 198 / 0.670 & 2005.9 & 319 & 1.47 & -11$\pm$14 & -32$\pm$2 & 262$\pm$15 & 53$\pm$15 \\
305.9110261 & 01 Apr 2021 & 198 / 0.949 & 2005.9 & 367 & 1.20 & -38$\pm$11 & 20$\pm$2 & 259$\pm$15 & 59$\pm$15 \\
326.8395710 & 22 Apr 2021 & 204 / 0.753 & 2005.9 & 279 & 1.74 & -64$\pm$15 & 17$\pm$2 & 212$\pm$15 & 37$\pm$15 \\
327.8120414 & 23 Apr 2021 & 205 / 0.023 & 2005.9 & 383 & 1.22 & -75$\pm$11 & 55$\pm$2 & 304$\pm$15 & 55$\pm$15 \\
328.8393031 & 24 Apr 2021 & 205 / 0.308 & 2005.9 & 362 & 1.28 & -76$\pm$12 & -63$\pm$2 & 369$\pm$15 & 69$\pm$15 \\
329.8575315 & 25 Apr 2021 & 205 / 0.590 & 2005.9 & 347 & 1.34 & -54$\pm$12 & -25$\pm$2 & 238$\pm$15 & 42$\pm$15 \\
330.8156155 & 26 Apr 2021 & 205 / 0.856 & 2005.9 & 286 & 1.75 & -78$\pm$15 & 22$\pm$2 & 148$\pm$15 & 26$\pm$15 \\
330.8364802 & 26 Apr 2021 & 205 / 0.861 & 2005.9 & 257 & 2.36 & -37$\pm$20 & & 150$\pm$15 & 24$\pm$15 \\
331.8534125 & 27 Apr 2021 & 206 / 0.143 & 2005.9 & 343 & 1.37 & -83$\pm$12 & 5$\pm$2 & 292$\pm$15 & 71$\pm$15 \\
332.8680461 & 28 Apr 2021 & 206 / 0.425 & 2005.9 & 318 & 1.61 & -72$\pm$14 & -48$\pm$2 & 183$\pm$15 & 28$\pm$15 \\
334.8263125 & 30 Apr 2021 & 206 / 0.968 & 2005.9 & 398 & 1.19 & -58$\pm$11 & 50$\pm$2 & 244$\pm$15 & 38$\pm$15 \\
335.8083082 & 01 May 2021 & 207 / 0.240 & 2005.9 & 379 & 1.22 & -75$\pm$12 & -52$\pm$2 & 403$\pm$15 & 82$\pm$15 \\
336.8470324 & 02 May 2021 & 207 / 0.528 & 2005.9 & 293 & 1.64 & -59$\pm$15 & -55$\pm$2 & 309$\pm$15 & 55$\pm$15 \\
\hline
\end{tabular} 
\label{tab:log}
\end{table*}

\setcounter{table}{0}
\begin{table*}
\caption[]{continued}
\begin{tabular}{cccccccccc}
\hline
BJD        & UT date & c / $\phi$ & t$_{\rm exp}$ & SNR & $\sigma_V$            & \Bl\   &  RV    & EW \pab\ & EW \brg\ \\ 
(2459000+) &         &            &   (s)        & ($H$) & ($10^{-4} I_c$)       & (G)   & (\ms)  & (\kms)   & (\kms)   \\  
\hline
649.8912529 & 11 Mar 2022 & 294 / 0.340 & 2005.9 & 401 & 1.19 & 17$\pm$11 & -24$\pm$2 & 406$\pm$15 & 73$\pm$15 \\
650.9038460 & 12 Mar 2022 & 294 / 0.621 & 2005.9 & 380 & 1.24 & 5$\pm$12 & 20$\pm$2 & 338$\pm$15 & 66$\pm$15 \\
651.8779503 & 13 Mar 2022 & 294 / 0.891 & 2005.9 & 410 & 1.16 & 22$\pm$11 & 34$\pm$2 & 393$\pm$15 & 87$\pm$15 \\
652.8769485 & 14 Mar 2022 & 295 / 0.168 & 2005.9 & 398 & 1.18 & 15$\pm$10 & -17$\pm$2 & 337$\pm$15 & 77$\pm$15 \\
653.9841063 & 15 Mar 2022 & 295 / 0.475 & 2005.9 & 389 & 1.20 & -16$\pm$10 & -44$\pm$2 & 366$\pm$15 & 76$\pm$15 \\
654.9433060 & 16 Mar 2022 & 295 / 0.741 & 1002.9 & 248 & 1.88 & 24$\pm$16 & -2$\pm$2 & 387$\pm$15 & 86$\pm$15 \\
655.9057570 & 17 Mar 2022 & 296 / 0.008 & 2005.9 & 328 & 1.76 & 25$\pm$16 & 4$\pm$2 & 300$\pm$15 & 61$\pm$15 \\
657.9321715 & 19 Mar 2022 & 296 / 0.570 & 2005.9 & 396 & 1.19 & -1$\pm$10 & -14$\pm$2 & 214$\pm$15 & 41$\pm$15 \\
658.9416328 & 20 Mar 2022 & 296 / 0.850 & 2005.9 & 353 & 1.36 & 16$\pm$12 & 39$\pm$2 & 242$\pm$15 & 43$\pm$15 \\
659.9525078 & 21 Mar 2022 & 297 / 0.130 & 2005.9 & 288 & 1.70 & 21$\pm$16 & 0$\pm$2 & 358$\pm$15 & 75$\pm$15 \\
660.9330715 & 22 Mar 2022 & 297 / 0.402 & 2005.9 & 313 & 1.56 & 37$\pm$13 & -16$\pm$2 & 343$\pm$15 & 64$\pm$15 \\
661.8884647 & 23 Mar 2022 & 297 / 0.667 & 2005.9 & 188 & 3.13 & -1$\pm$28 & & 243$\pm$15 & 46$\pm$15 \\
661.9123263 & 23 Mar 2022 & 297 / 0.674 & 2005.9 & 231 & 2.43 & 9$\pm$22 & 2$\pm$2 & 240$\pm$15 & 43$\pm$15 \\
678.9220651 & 09 Apr 2022 & 302 / 0.391 & 2005.9 & 360 & 1.41 & 7$\pm$13 & -13$\pm$2 & 244$\pm$15 & 46$\pm$15 \\
681.8732563 & 12 Apr 2022 & 303 / 0.209 & 2005.9 & 293 & 1.65 & 14$\pm$16 & -35$\pm$2 & 307$\pm$15 & 52$\pm$15 \\
682.8548885 & 13 Apr 2022 & 303 / 0.482 & 2005.9 & 332 & 1.41 & -1$\pm$13 & 0$\pm$2 & 400$\pm$15 & 73$\pm$15 \\
683.8634512 & 14 Apr 2022 & 303 / 0.761 & 2005.9 & 350 & 1.35 & 28$\pm$13 & 14$\pm$2 & 300$\pm$15 & 52$\pm$15 \\
684.7962789 & 15 Apr 2022 & 304 / 0.020 & 2005.9 & 365 & 1.30 & 2$\pm$12 & -17$\pm$2 & 276$\pm$15 & 49$\pm$15 \\
687.8950037 & 18 Apr 2022 & 304 / 0.879 & 2005.9 & 344 & 1.41 & 29$\pm$15 & 33$\pm$2 & 318$\pm$15 & 52$\pm$15 \\
690.8404521 & 21 Apr 2022 & 305 / 0.696 & 2005.9 & 356 & 1.37 & 15$\pm$15 & 33$\pm$2 & 265$\pm$15 & 51$\pm$15 \\
710.8180850 & 11 May 2022 & 311 / 0.236 & 2005.9 & 263 & 1.96 & 30$\pm$19 & -17$\pm$2 & 240$\pm$15 & 40$\pm$15 \\
711.8161622 & 12 May 2022 & 311 / 0.513 & 2005.9 & 383 & 1.24 & 1$\pm$11 & 38$\pm$2 & 277$\pm$15 & 61$\pm$15 \\
712.8124998 & 13 May 2022 & 311 / 0.789 & 2005.9 & 300 & 1.62 & 24$\pm$15 & 14$\pm$2 & 245$\pm$15 & 42$\pm$15 \\
713.7983426 & 14 May 2022 & 312 / 0.063 & 2005.9 & 300 & 1.61 & 39$\pm$16 & -34$\pm$2 & 350$\pm$15 & 81$\pm$15 \\
714.7779533 & 15 May 2022 & 312 / 0.334 & 2005.9 & 349 & 1.38 & 37$\pm$13 & -20$\pm$2 & 300$\pm$15 & 53$\pm$15 \\
715.8090560 & 16 May 2022 & 312 / 0.620 & 2005.9 & 338 & 1.42 & 16$\pm$13 & 8$\pm$2 & 242$\pm$15 & 45$\pm$15 \\
716.8108812 & 17 May 2022 & 312 / 0.898 & 2005.9 & 385 & 1.24 & 21$\pm$12 & -13$\pm$2 & 306$\pm$15 & 65$\pm$15 \\
719.8653011 & 20 May 2022 & 313 / 0.745 & 2005.9 & 342 & 1.44 & -2$\pm$15 & -4$\pm$3 & 209$\pm$15 & 34$\pm$15 \\
\hline
\end{tabular} 
\end{table*}

\section{Stacked periodogram of TESS light curves} 
\label{sec:appT}

We show in this section the stacked periodograms of the TESS light curves of TW~Hya collected in March 2019 and 2021.  

\begin{figure*}
\centerline{\includegraphics[scale=0.5,angle=-90]{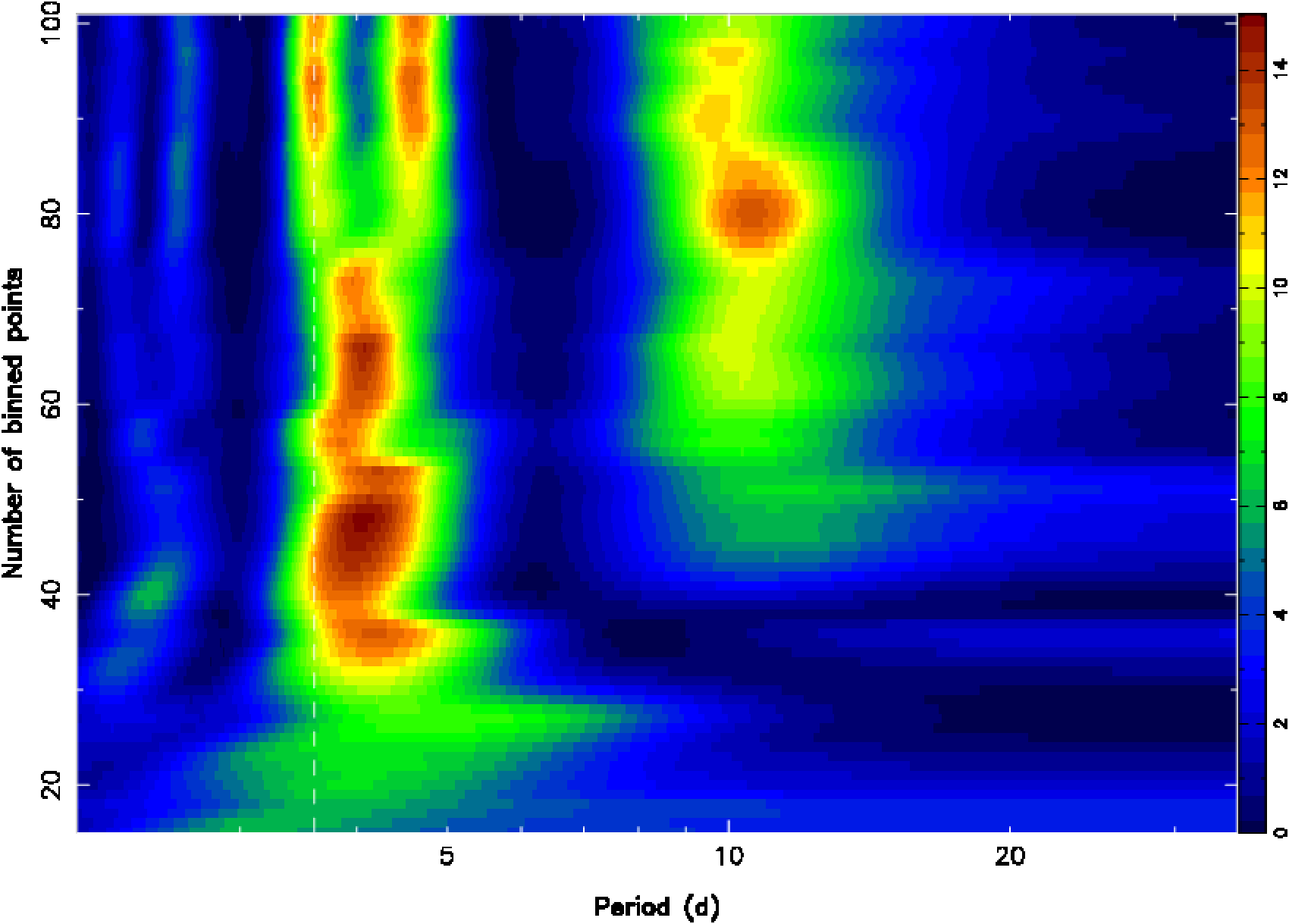}\vspace{5mm}} 
\centerline{\includegraphics[scale=0.5,angle=-90]{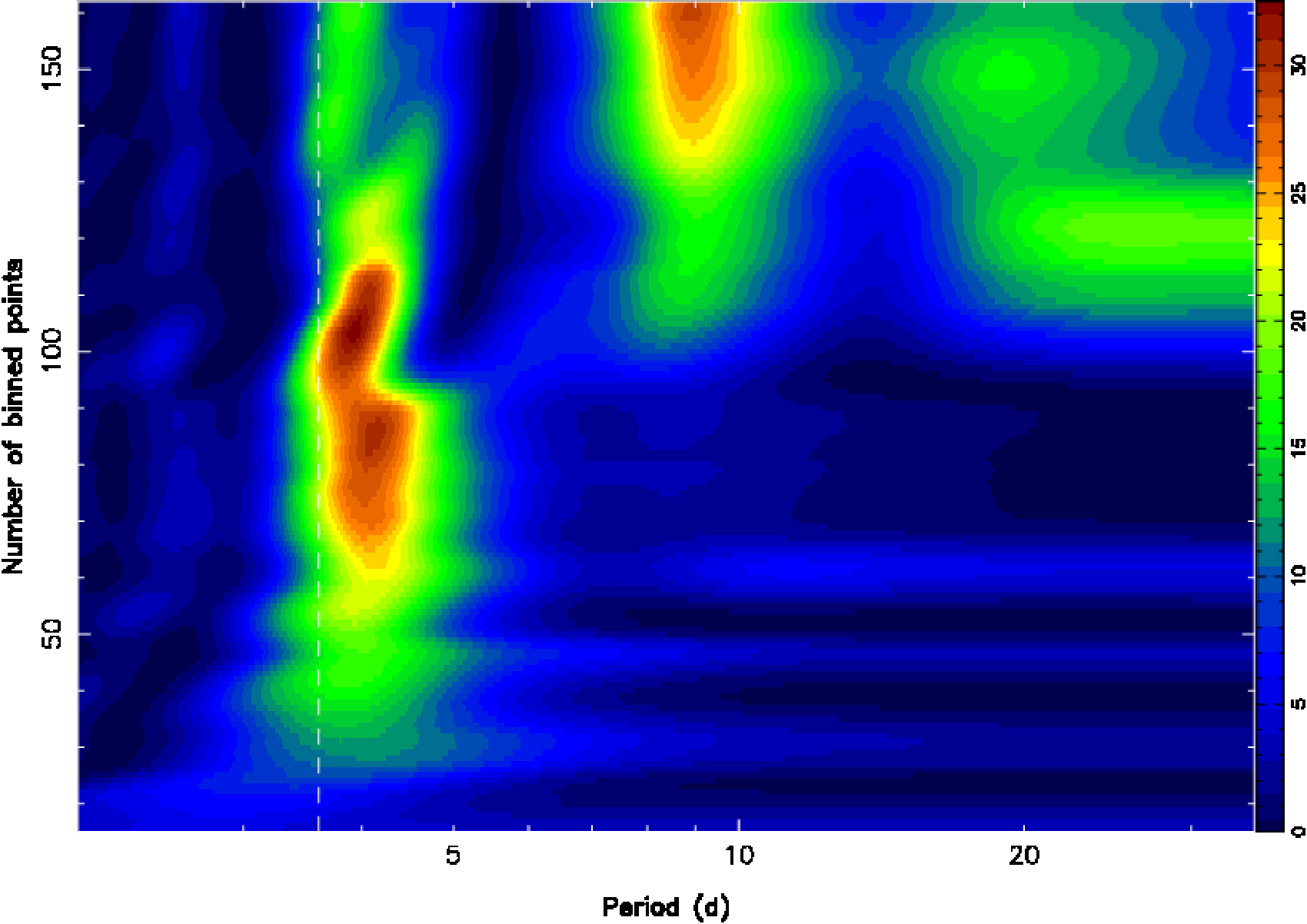}} 
\caption[]{Stacked periodograms of the TESS light curves of TW~Hya collected in March 2019 (top) and 2021 (bottom), {\emr with the TESS data binned 
by groups of 10 (in 2019) and 20 (in 2021) adjacent points.  In these plots, each horizontal line corresponds to a color-coded periodogram, computed on 
an increasing number of binned points (starting from the first binned point).  The color scale depicts the logarithmic power of the periodogram.  }
At both epochs, a signal at a period of about 4~d shows up, but only dominates for a limited time before splitting itself into 2 weaker signals (one 
of which at the rotation period, in 2019) or vanishing (in 2021).  In 2021 (and 2023, not shown), power at a period of about 9~d dominates the overall 
light curve, but not in 2019 where power at about 10~d quickly weakens after showing up.  The dashed line depicts the rotation period.  }  
\label{fig:tes}
\end{figure*}

\section{Detailed results of the PCA analysis} 
\label{sec:appP}

We present here the full results of the PCA analysis applied to our Stokes $V$ profiles of TW~Hya.  

\begin{figure*}
        \raggedright \hspace{3.0cm} \textbf{a.} \hspace{4.5cm} \textbf{b.} \\
        \centerline{\bf\large all data \raisebox{-0.8\totalheight}{
        \includegraphics[width=0.55\columnwidth, bb=0 0 505.246875 412.25875,trim={0 0 0 0}, clip]{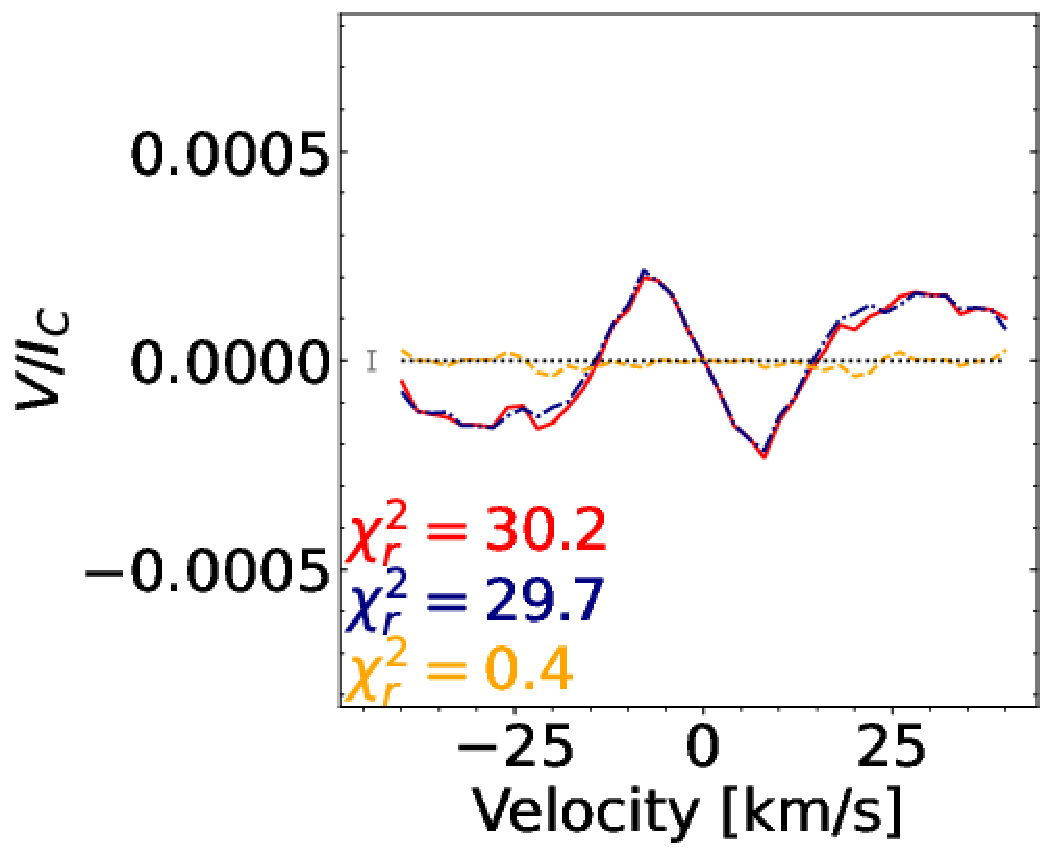} \hspace{2mm}
        \includegraphics[width=\columnwidth, bb=0 0 1305.009375 781.61875, trim={0 400 440 0}, clip]{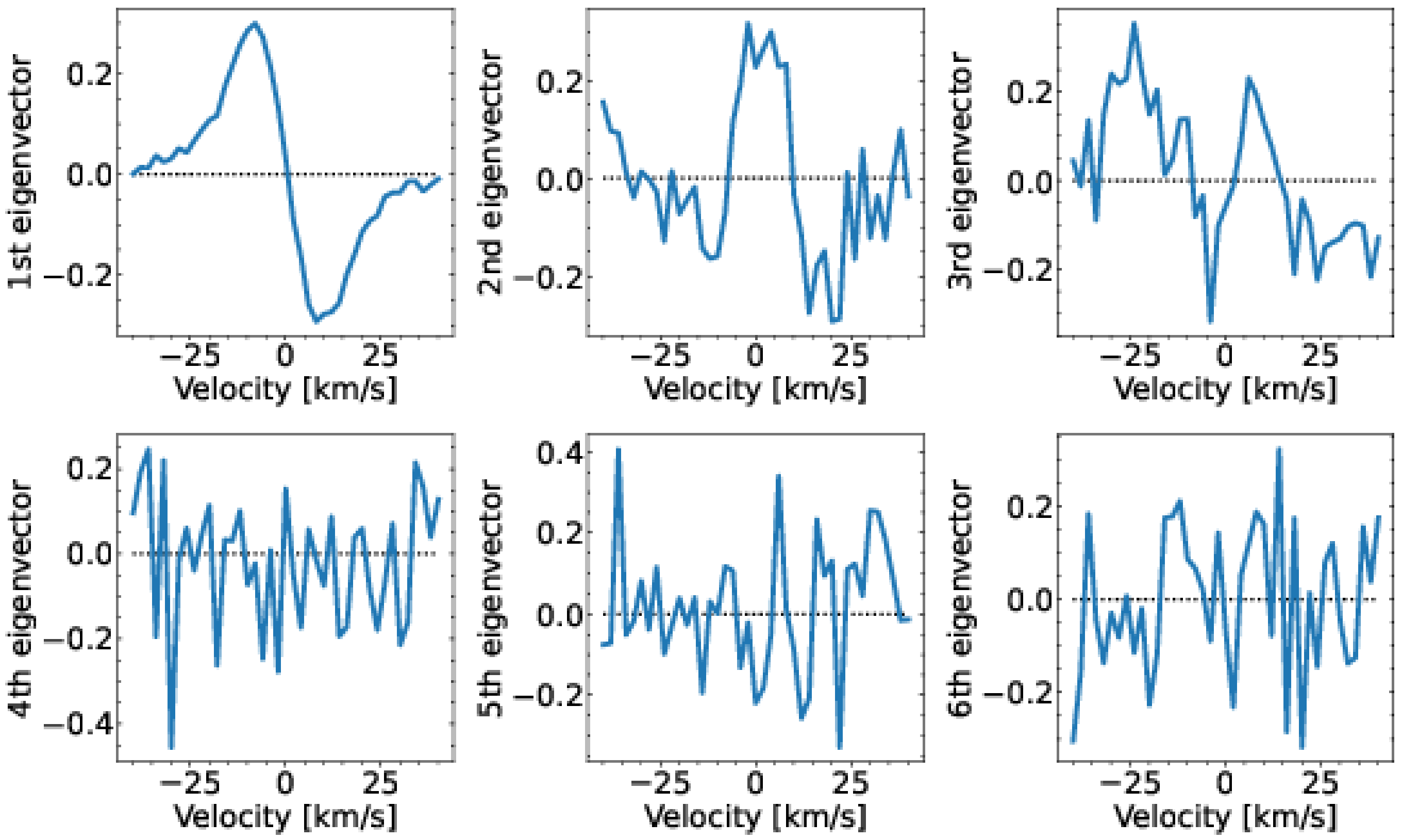}}}
        
\vspace{2mm}
        \raggedright \hspace{3.0cm} \textbf{c.} \\
        \centerline{\bf\large 2019 \raisebox{-0.8\totalheight}{
        \includegraphics[width=0.55\columnwidth, bb=0 0 505.246875 401.17075, trim={0 0 0 0}, clip]{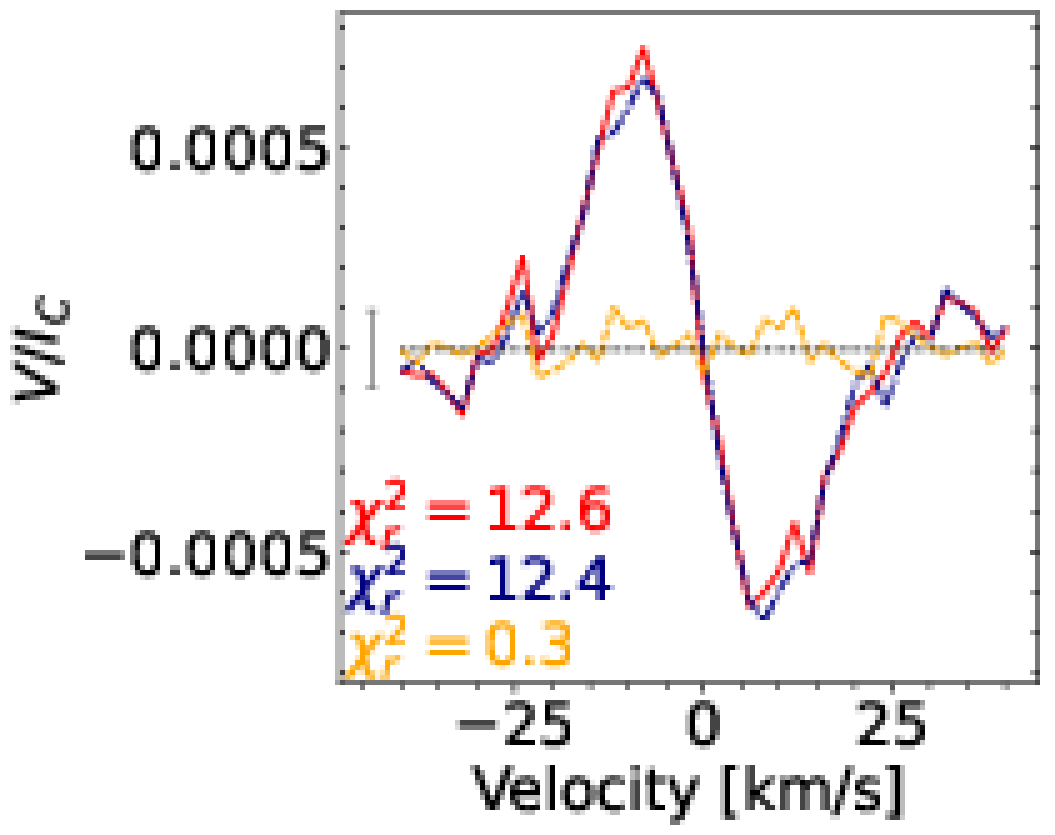}
        \includegraphics[width=\columnwidth, bb=0 0 1340.634375 781.61875, trim={0 400 450 0}, clip]{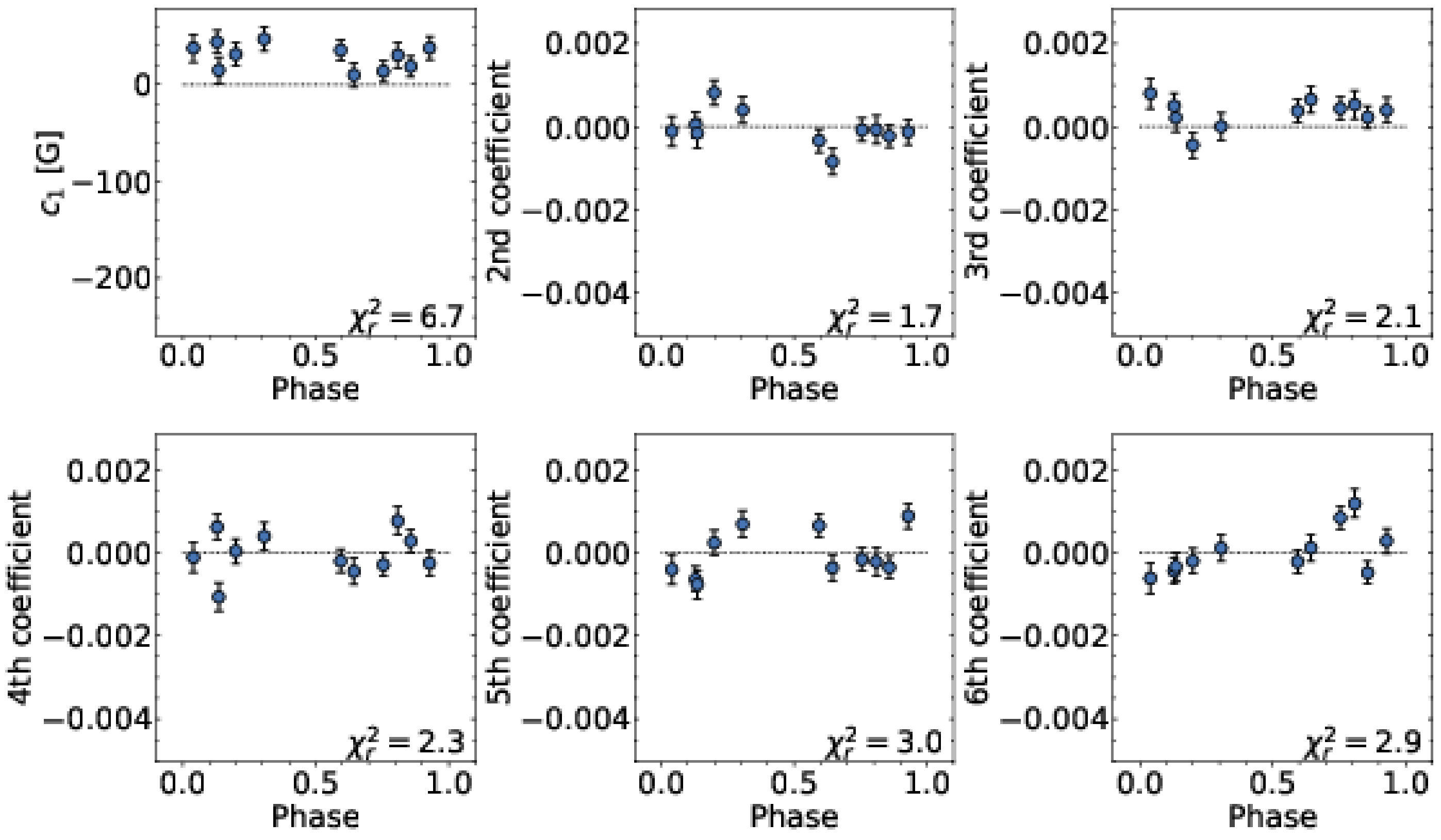}}} 
       
        \centerline{\bf\large 2020 \raisebox{-0.8\totalheight}{
        \includegraphics[width=0.55\columnwidth, bb=0 0 505.246875 401.17075, trim={0 0 0 0}, clip]{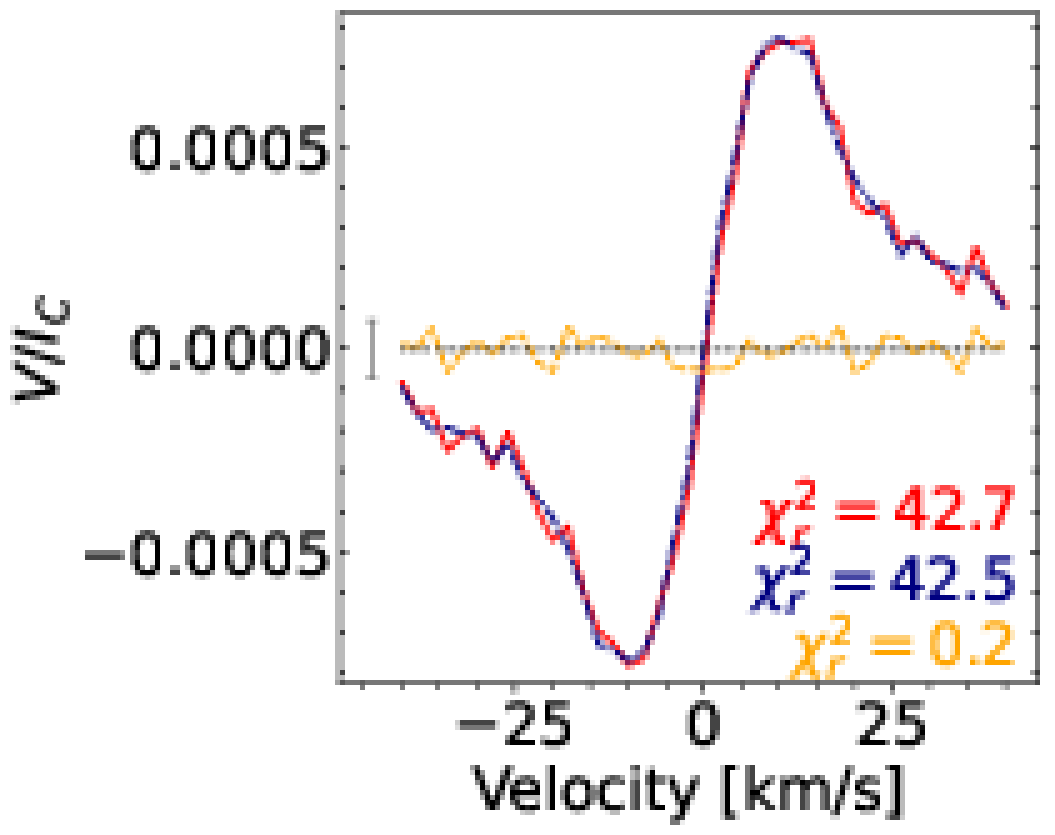}
        \includegraphics[width=\columnwidth, bb=0 0 1340.634375 781.61875, trim={0 400 450 0}, clip]{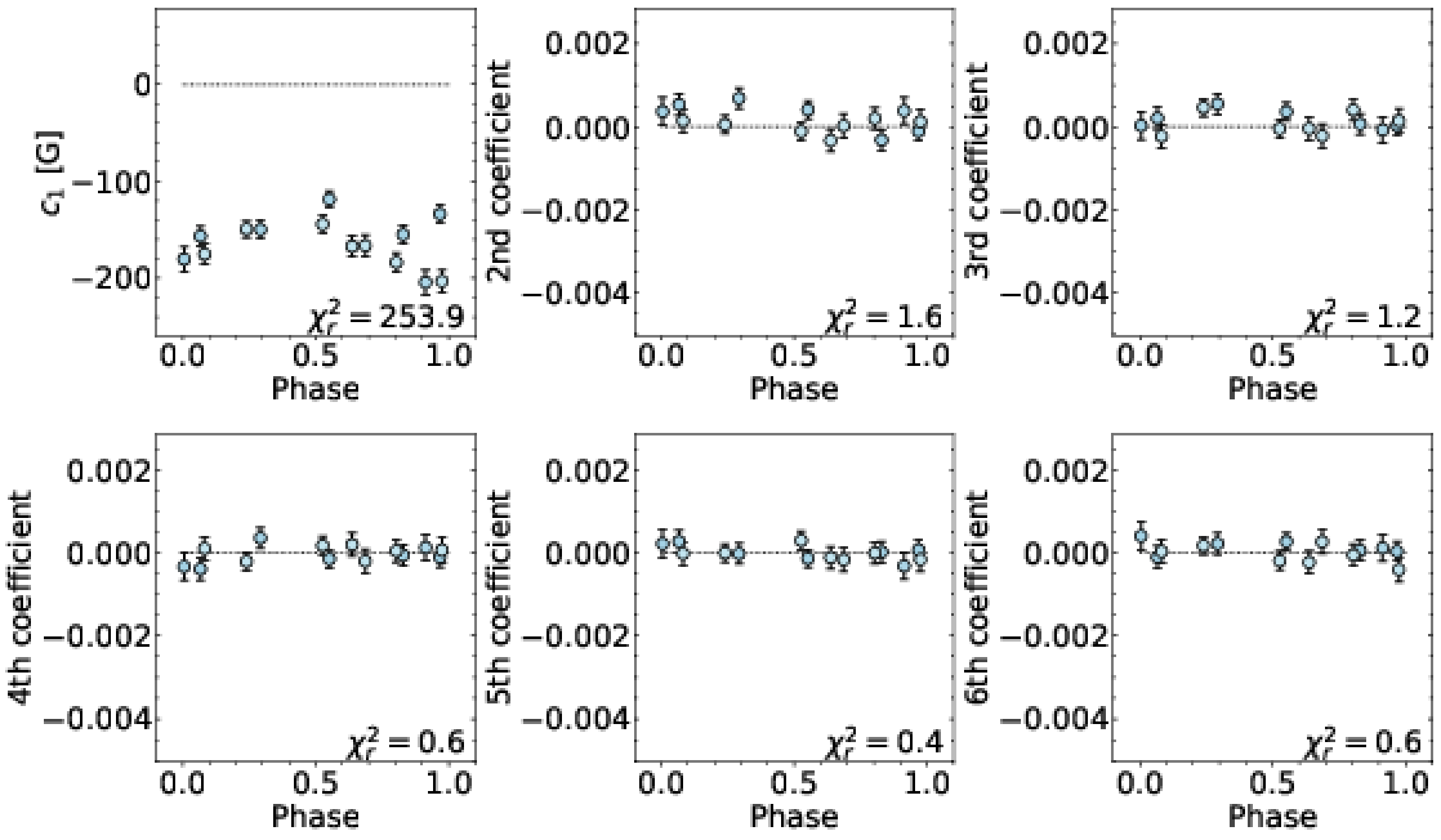}}}
       
        \centerline{\bf\large 2021 \raisebox{-0.8\totalheight}{
        \includegraphics[width=0.55\columnwidth, bb=0 0 505.246875 401.17075, trim={0 0 0 0}, clip]{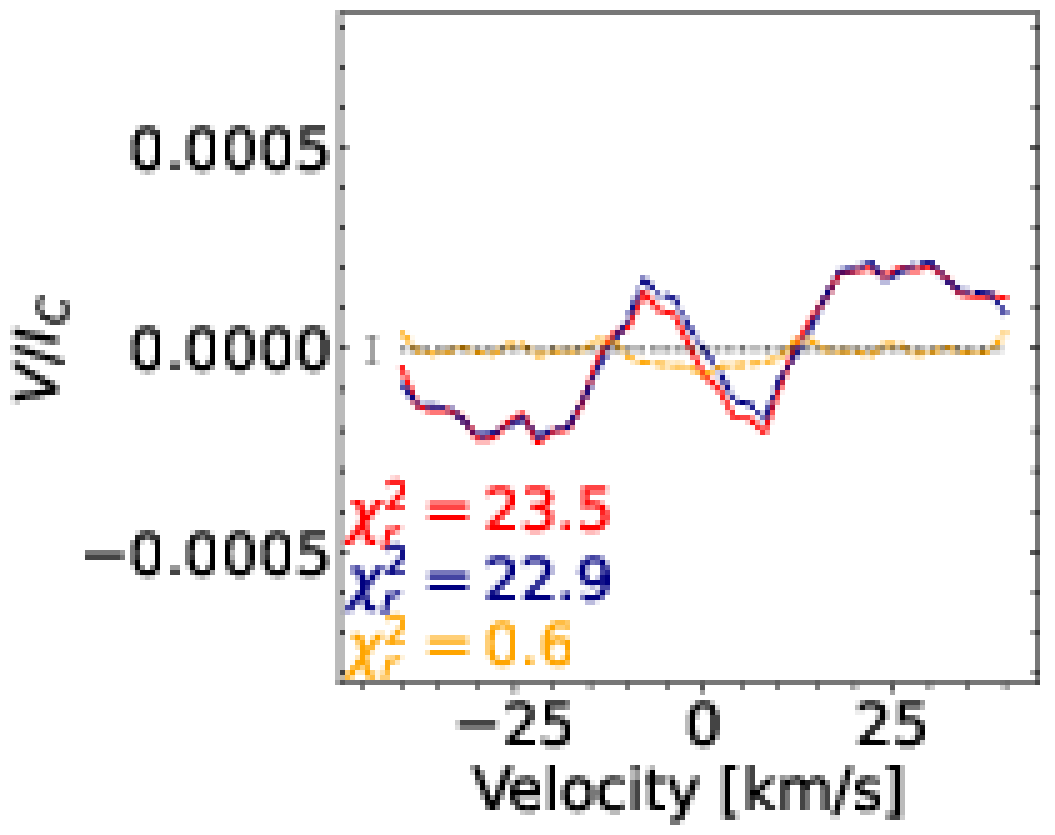}
        \includegraphics[width=\columnwidth, bb=0 0 1340.634375 781.61875, trim={0 400 450 0}, clip]{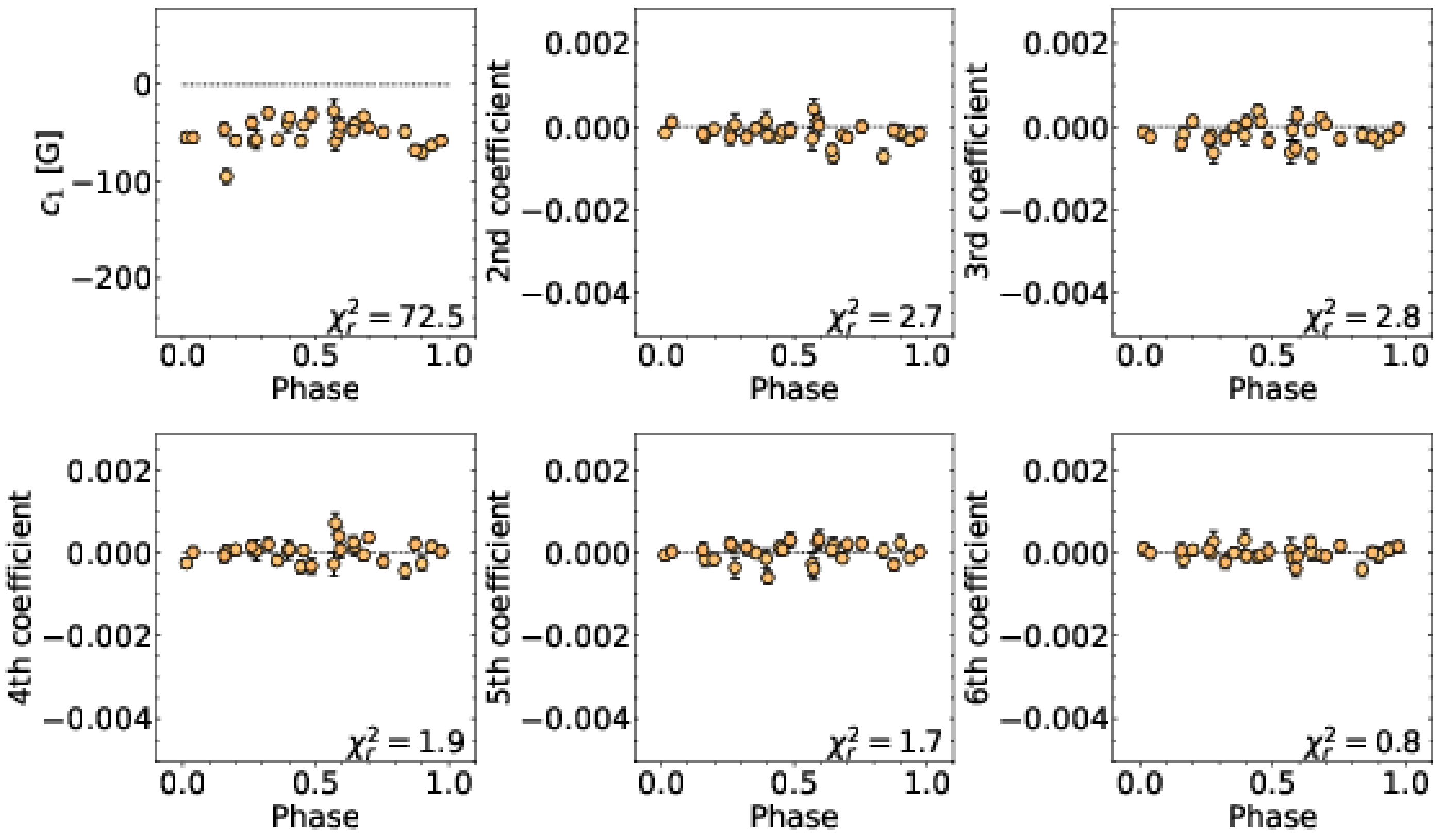}}}
       
        \centerline{\bf\large 2022 \raisebox{-0.8\totalheight}{
        \includegraphics[width=0.55\columnwidth, bb=0 0 505.246875 401.17075, trim={0 0 0 0}, clip]{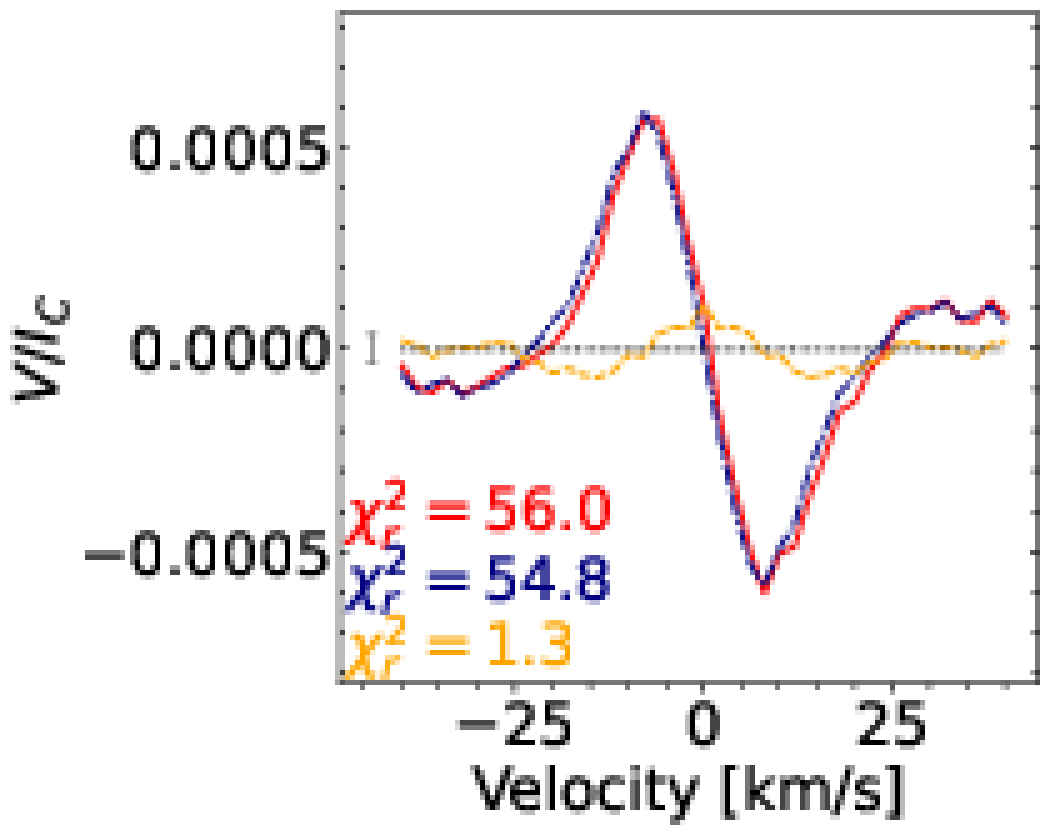}
        \includegraphics[width=\columnwidth, bb=0 0 1340.634375 781.61875, trim={0 400 450 0}, clip]{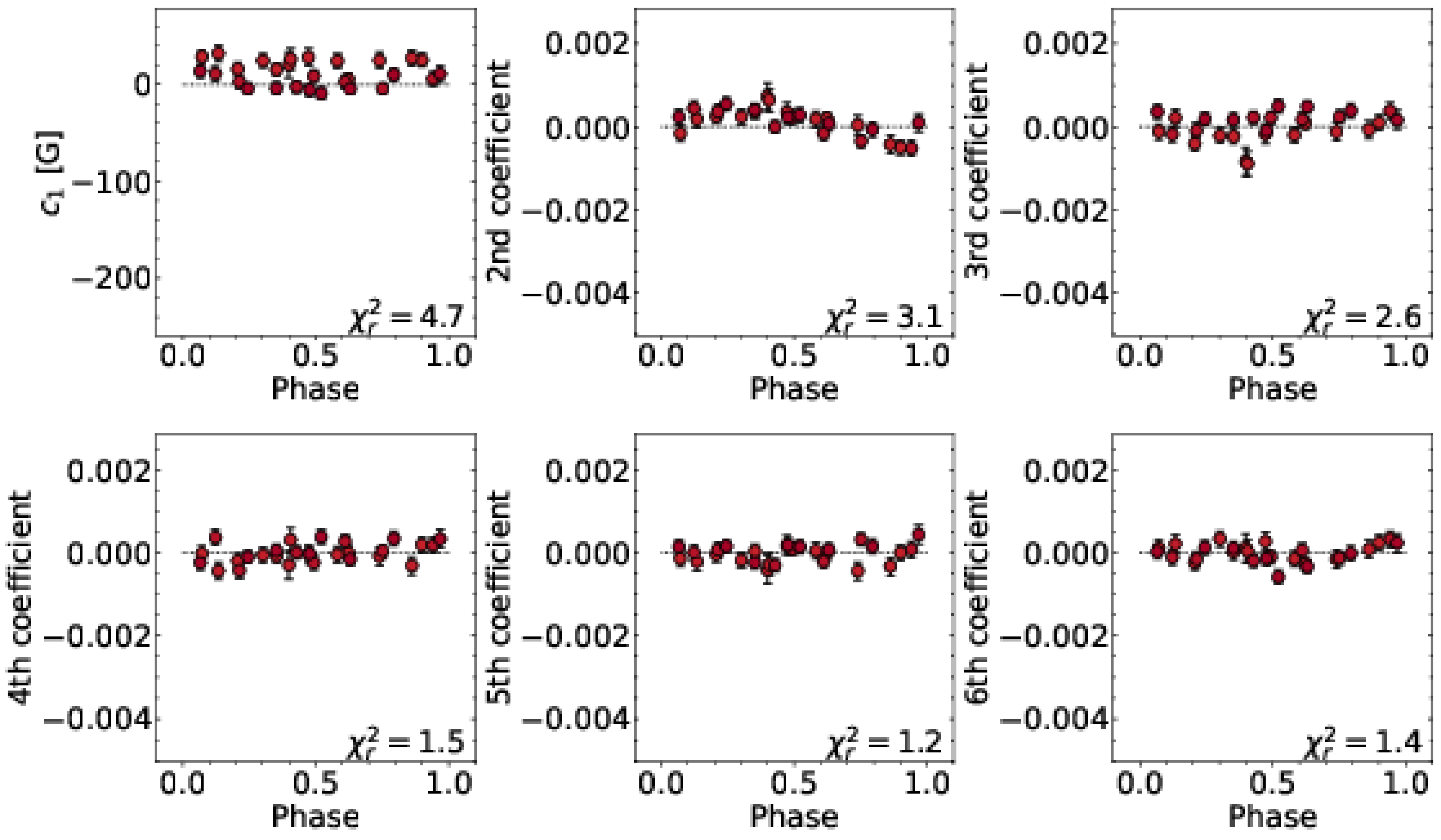}}} 
\caption{Detailed results of the PCA analysis on our Stokes $V$ profiles of TW~Hya. \textbf{a.} The weighted average of all observations (full red line), 
and its decomposition into the antisymmetric (blue dash-dotted line) and symmetric (orange dashed line) components with respect to the line centre, which relate to 
axisymmetric components of the poloidal and toroidal large-scale field of TW~Hya respectively \citep{Lehmann22}.  This mean profile is used to compute the mean-subtracted Stokes~$V$ 
profiles to which PCA is applied, with the derived eigenvectors and coefficients shown in panels b and c. \textbf{b.} The first two eigenvectors derived with PCA from the mean-subtracted 
Stokes~$V$ profiles. \textbf{c.} For each of our 4 observing seasons (one season per row), we show, from left to right, the mean profiles (as described in panel a), the 
first PCA coefficient $c_1$ (scaled and shifted to match \Bl) and the second PCA coefficient as a function of rotation phase. The \chisqr\ associated with each 
profile and coefficient time series is included in all relevant plots, showing that the toroidal component of the large-scale field is undetected ($\chisqr\simeq1$) 
whereas the second PCA coefficient is only marginally significant.  }
    \label{fig:pca}
\end{figure*}

\section{Complementary information on RV analysis} 
\label{sec:appB}

\begin{figure*}
\centerline{\includegraphics[scale=0.35,angle=-90]{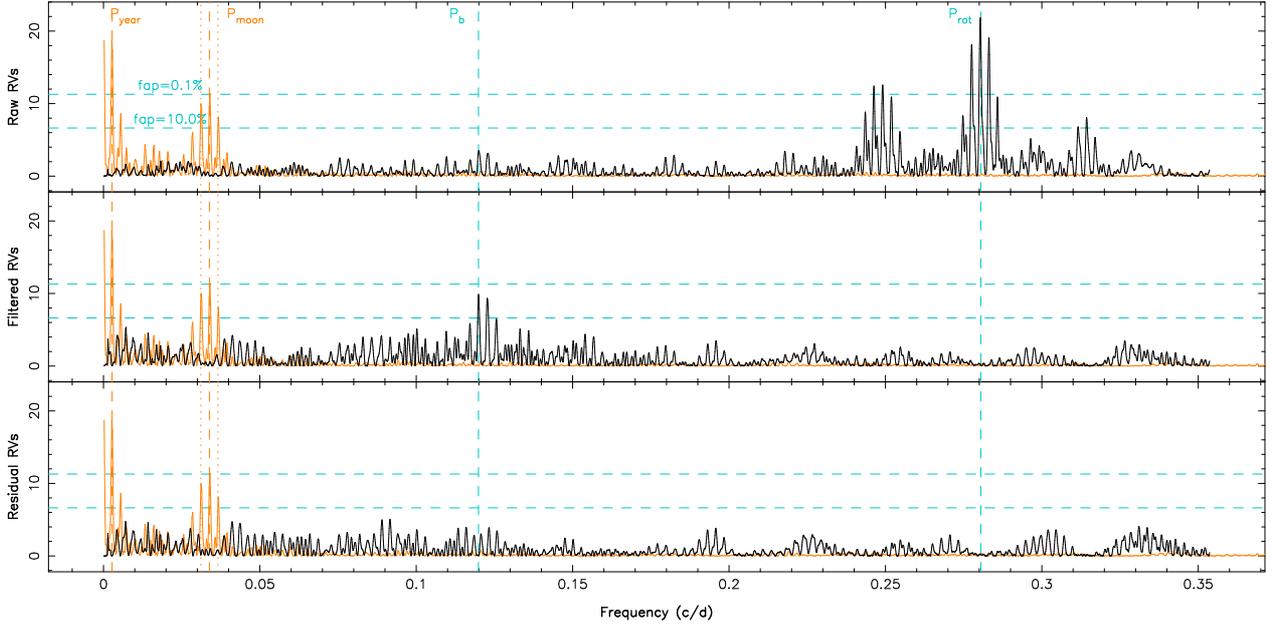}} 
\caption[]{Periodogram of the raw (top), filtered (middle) and residual (bottom) RVs when including a fit to the RV signal at a period of $P_b=8.34$~d 
in the MCMC modeling.  The cyan vertical dashed lines respectively trace \Prot\ and $P_b$, whereas the horizontal 
dashed lines indicate the 10 and 0.1\% FAP levels in the periodogram of our RV data.  The orange curve depicts the window function, whereas the orange 
vertical dashed and dotted line outline the 1-yr period, the synodic period of the Moon (at 29.5~d) and its 1-yr aliases.  }
\label{fig:per}
\end{figure*}

\begin{figure*}
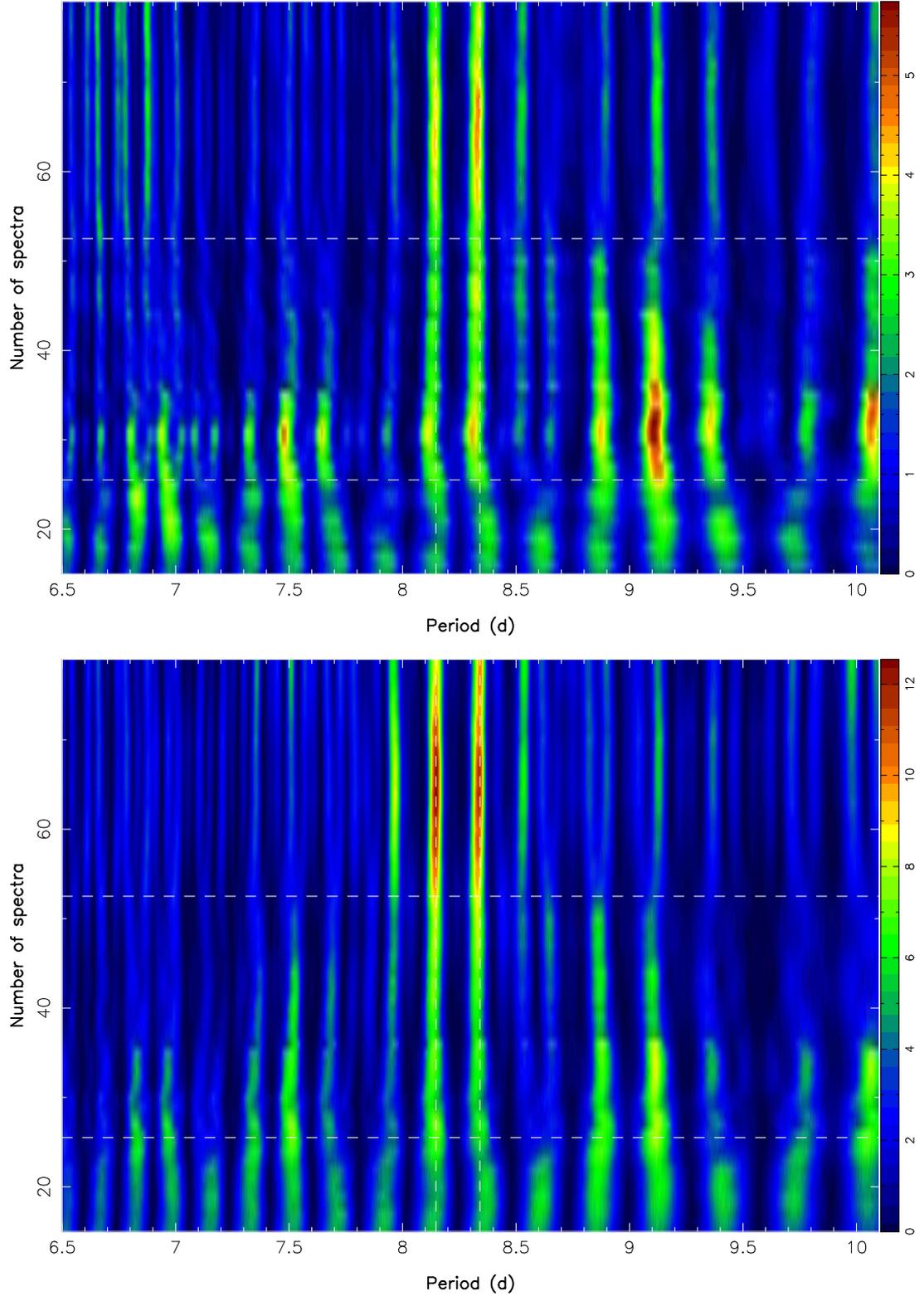

\centerline{\includegraphics[scale=0.5,angle=-90]{fig/twhya-str.ps}\vspace{3mm}} 
\centerline{\includegraphics[scale=0.5,angle=-90]{fig/twhya-stf.ps}} 
\caption[]{Stacked periodograms of the raw (top) and filtered (bottom) RVs, as a function of the number of successive spectra taken into account in 
the Fourier analysis (starting from the first collected spectrum).  The main RV signal and its main 1-yr alias (outlined with vertical dashed lines at 8.34~d 
and 8.15~d, see Table~\ref{tab:pla}) get stronger and increasingly dominant as more spectra are added to the analysis, especially in the filtered RVs 
but also in the raw RVs (though at a lower significance level as expected).  The horizontal dashed lines illustrate the transition between seasons 2020 and 2021, 
and between 2021 and 2022.  Note the difference in color scale between both panels.  }
\label{fig:stp}
\end{figure*}

\begin{figure*}
\centerline{\includegraphics[scale=1.0,angle=-90,bb=0 0 480 480]{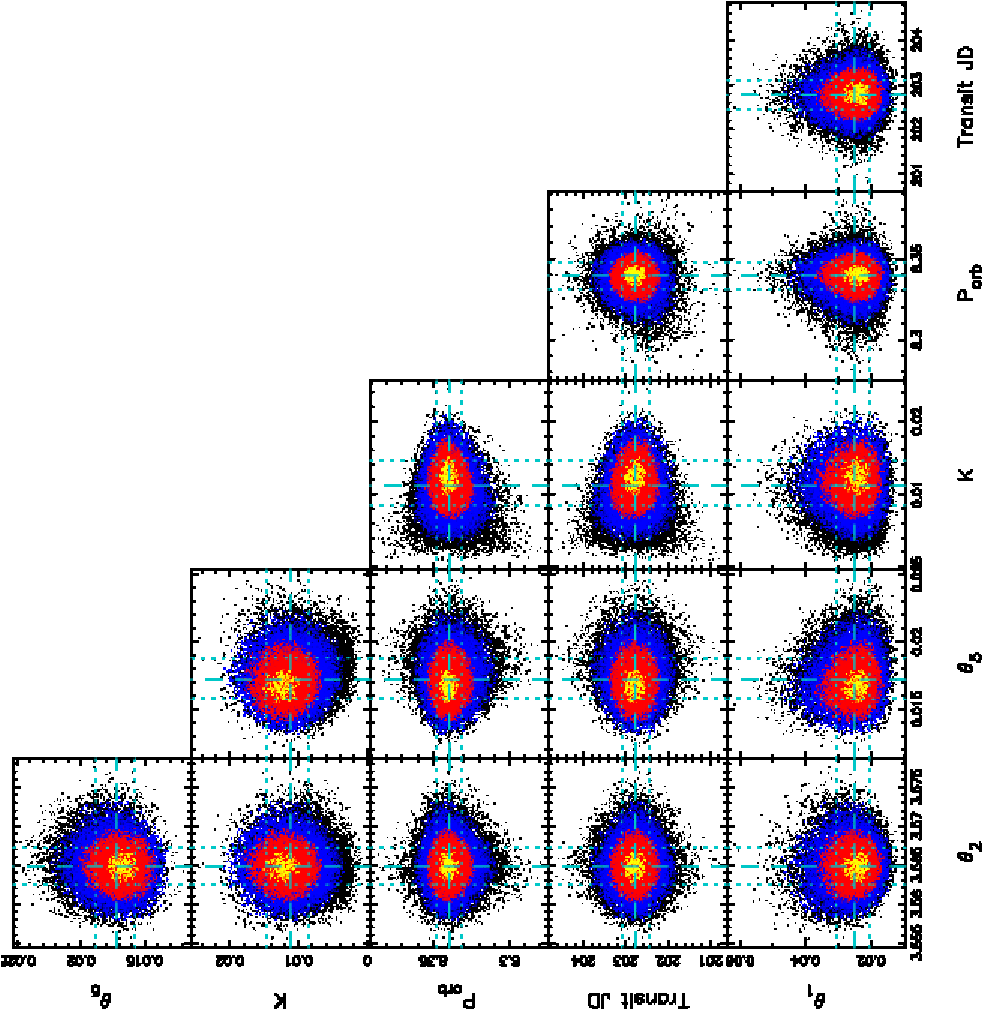}} 
\caption[]{Corner plot of the MCMC fit to our RV data.  The yellow, red and blue regions depict the 1, 2 and 3 $\sigma$ confidence regions, 
whereas the cyan dashed and dotted lines denote the optimal values and corresponding error bars. } 
\label{fig:cor}
\end{figure*}

We show in Fig.~\ref{fig:per} the periodogram of raw, filtered and residual RVs of TW~Hya over the full set of our observations, and in Fig.~\ref{fig:stp} 
the stacked periodograms for the raw and filtered RVs.  In Fig.~\ref{fig:cor}, we show the corner plot of our MCMC fit to the RV data.  All plots refer to 
case b of Table \ref{tab:pla}.  

\section{Profiles, 2D periodogram and Zeeman signature for \brg}
\label{sec:appC}

\begin{figure}
\centerline{\hspace{-2mm}\includegraphics[scale=0.3,angle=-90]{fig/twhya-brg.ps}\vspace{2mm}}
\centerline{\includegraphics[scale=0.55,angle=-90]{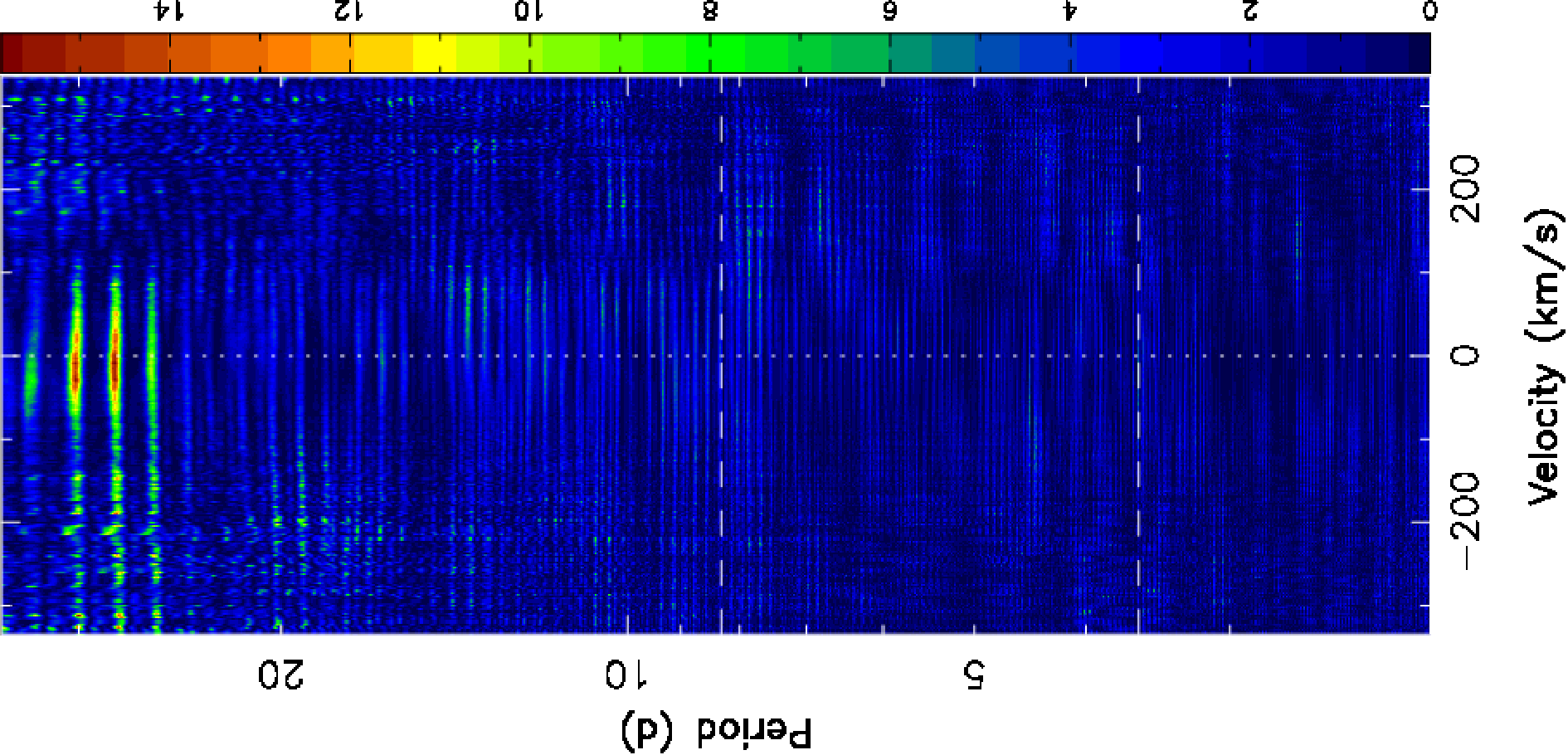}}
\caption[]{Same as Fig.~\ref{fig:eml} for \brg.}  
\label{fig:eml2}
\end{figure}

We show in Fig.~\ref{fig:eml2} the stacked profiles and 2D periodogram of \brg\ over the full set of our observations.  
Fig.~\ref{fig:emv2} depicts the average \brg\ profile and the associated Zeeman signature.  

\begin{figure}
\centerline{\includegraphics[scale=0.35,angle=-90]{fig/twhya-brgv.ps}}
\caption[]{Same as Fig.~\ref{fig:emv} for \brg.} 
\label{fig:emv2}
\end{figure}

\bsp    
\label{lastpage}
\end{document}